\documentclass[11pt]{article}
\usepackage{amsmath,amssymb}
\usepackage{hyperref}
\usepackage{graphicx}
\usepackage{eufrak}

\newcommand{\bea}{\begin{eqnarray}}
\newcommand{\eea}{\end{eqnarray}}

\newcommand{\ee}{\mathrm{e}}

\newsavebox{\uuunit}
\sbox{\uuunit}
    {\setlength{\unitlength}{0.825em}
     \begin{picture}(0.6,0.7)
        \thinlines
        \put(0,0){\line(1,0){0.5}}
        \put(0.15,0){\line(0,1){0.7}}
        \put(0.35,0){\line(0,1){0.8}}
       \multiput(0.3,0.8)(-0.04,-0.02){10}{\rule{0.5pt}{0.5pt}}
     \end {picture}}

\numberwithin{equation}{section}

\begin{document}
\begin{titlepage}
\begin{center}
\hfill LMU-ASC 11/08 \\
\hfill MPP-2008-17 \\
\vskip 6mm

{\Large \textbf{
On five-dimensional non-extremal charged 
black holes  \\
\vskip 1mm
and FRW cosmology
}}
\vskip 8mm

\textbf{G. L.~Cardoso$^{a}$ and V.~Grass$^{a,b}$}

\vskip 4mm
$^a${\em Arnold Sommerfeld Center for Theoretical Physics\\
Department f\"ur Physik,
Ludwig-Maximilians-Universit\"at M\"unchen \\ Theresienstr. 37, 
80333 M\"unchen, Germany}
\\[1mm]
$^b${\em Max-Planck-Institut f\"ur Physik\\
F\"ohringer Ring 6, 80805 M\"unchen, Germany
}\\[1mm]

\centerline{
{\tt gcardoso,viviane@theorie.physik.uni-muenchen.de}}

\end{center}
\vskip .2in
\begin{center} {\bf ABSTRACT } \end{center}
\begin{quotation}\noindent

We consider static non-extremal charged black hole solutions
in the context of $N=2$
gauged supergravity theories in five dimensions, and
we show that they satisfy first-order flow
equations.  Then we analyze the motion of the dual
brane in these black hole backgrounds.
We express the entropy in terms of 
a Cardy-Verlinde-type formula, and
we show that the equations describing the FRW cosmology on the brane
have a form that is similar to the equations 
for the entropy and for the Casimir energy on the brane.
We also briefly comment on the inclusion of
a Gauss-Bonnet term in the analysis.

\end{quotation}

\vfill
\end{titlepage}
\eject


\section{Introduction}

\setcounter{equation}{0}

In \cite{Verlinde:2000wg} a surprising relation was found
between the Friedmann-Robertson-Walker (FRW) equations of standard
cosmology describing a universe filled with radiation and the 
equations describing the entropy and the Casimir
energy of the radiation.  The connection came about by representing
the radiation by a strongly coupled conformal field theory (CFT) with
an AdS-dual description in terms of a
five-dimensional  AdS-Schwarzschild black hole \cite{Witten:1998zw}.
This connection was further studied in \cite{Savonije:2001nd} from a 
brane-world perspective \cite{Randall:1999ee,Randall:1999vf},
where the universe was represented in terms of
a spherical brane moving in the black hole background \cite{Gubser:1999vj}.  
The universe, which is described by the standard FRW 
equations, has a size that corresponds to the distance
of the brane to the black hole singularity.
The brane first expands until it reaches a maximal
radius after which it recontracts and falls through the horizon.
It was found \cite{Verlinde:2000wg} that the entropy of the radiation
can be written in a form analogous to Cardy's formula for the entropy 
of a two-dimensional CFT \cite{Cardy:1986ie}. 
This is the so-called Cardy-Verlinde
formula.
Moreover, 
the Cardy-Verlinde formula and the equation for the Casimir energy 
have a form that is similar to the two 
FRW equations on the brane, and these two sets of equations 
coincide 
as the brane crosses the horizon of the black hole.
This may indicate
\cite{Verlinde:2000wg,Savonije:2001nd} that
the standard FRW equations and the equations for the entropy and the energy
of the CFT
have a common origin in string or M-theory.

The Cardy-Verlinde formula for the radiation CFT mentioned above expresses
the square of the entropy ${\cal S}$ in terms of 
the product of the extensive  
part $E_e$ of the energy  and the Casimir energy $E_c$ on the brane.
The Casimir energy is defined as the violation of the 
thermodynamic Euler relation \cite{Verlinde:2000wg}.  
The energies $E_e$ and $E_c$ behave as 
\begin{equation}
E_e \, a \propto {\cal S}^{4/3} \;\;\;,\;\;\; E_c \,  a 
\propto {\cal S}^{2/3} \;,
\label{scaleeec}
\end{equation}
where $a$ denotes the radius of the spherical brane.
The proportionality coefficients in \eqref{scaleeec} are independent
of $a$ and ${\cal S}$.
In the following we consider 
charged  AdS black holes in 
five-dimensional
$N=2$ gauged supergravity theories
obtained by gauging 
the $U(1)$ subgroup of the $SU(2)$-automorphism group of
the $N=2$ supersymmetry algebra \cite{Gunaydin:1984ak}. 
These 
theories have a potential which is
constructed out of a 
superpotential $W$ that depends on the scalar fields of the theory.
It turns out to be convenient to also introduce a `dual' superpotential
${\tilde W}$. 
We determine two quantities that are of the form \eqref{scaleeec} and that
we denote by $E_e$ and $E_c$.  The latter is the Casimir energy of
the field theory on the brane dual to the charged 
AdS black hole.
They come
with proportionality
coefficients given by $W$ and ${\tilde W}$
evaluated at the event horizon of the black hole, respectively.
We note, however, that 
$W$ and  ${\tilde W}$ do not have a simple 
dependence on extensive quantities.
We express the entropy as 
a Cardy-Verlinde-type formula in terms of the product of $E_e$ and $E_c$.
For a discussion 
of the Cardy-Verlinde formula in the
context of the STU model see also  \cite{Cai:2001jc,Klemm:2001pn}.
We show that 
the two FRW equations describing the motion of the brane
in the charged black hole background
take a form that is similar to 
the Cardy-Verlinde-type formula and to the equation for 
the Casimir energy on the brane, respectively.
Then, as the brane crosses the event horizon of the black hole, 
these two sets of equations 
again coincide.
This has already been addressed in 
\cite{Biswas:2001sh,Cai:2001ur,Youm:2001qr,Gregory:2002am}
for the case of a static AdS-Einstein-Maxwell black hole.
We also discuss the Bekenstein bound 
on the entropy \cite{Bekenstein:1980jp,Verlinde:2000wg}
for some of the black hole solutions arising in the STU model.

The five-dimensional 
black hole solutions that we consider in the context of the $N=2$
gauged supergravity theories mentioned above are 
the 
non-extremal electrically 
charged static AdS black holes 
of 
\cite{Behrndt:1998jd}, which were constructed
by solving the associated equations of motion.
Special cases of these black holes have
been discussed in \cite{Chamblin:1999tk,Cvetic:1999ne,Chamblin:1999hg,
Hawking:1999dp,Gubser:2004xx,
Gao:2004tv,Elvang:2007ba,Astefanesei:2007vh}.
We rederive the black hole solutions of 
\cite{Behrndt:1998jd} by showing that they satisfy first-order flow
equations.  These are obtained by rewriting the five-dimensional bulk
action in terms of squares of first-order differential equations.
Since the black hole solutions are non-extremal, there are additional terms in
the action which are not of the square type, but which are consistent
with the first-order flow equations.  First-order flow equations
for non-extremal 
AdS-Einstein-Maxwell black holes have been discussed in the past in 
\cite{Lu:2003iv} and for extremal AdS black holes supported
by one scalar field in \cite{Elvang:2007ba}.
More recently, such equations have also been 
discussed in 
\cite{Miller:2006ay,Janssen:2007rc}
in the context of 
Einstein-dilaton-p-form systems.

The paper is organized as follows.  In section 2 we review
a few relevant facts about the  five-dimensional
$N=2$ gauged supergravity theories that we will be looking at.  Then
we show that the black hole solutions of 
\cite{Behrndt:1998jd} satisfy first-order flow
equations.  We briefly 
discuss various black hole solutions in the context of the
STU model.  Then we give the gravitational counterpart of 
the energies $E_e$ and $E_c$
at the moment when the brane crosses the event horizon of the black hole.
We denote these by ${\tilde E}_e$ and ${\tilde E_c}$ respectively,
and we discuss their relation to the 
Smarr formula \cite{Gauntlett:1998fz} for charged black holes. 
The quantities ${\tilde E}_e$ and ${\tilde E_c}$ are proportional
to $W$ and ${\tilde W}$ evaluated at the horizon, respectively.
The resulting Cardy-Verlinde-type formula, which is expressed 
in terms of 
${\tilde E}_e \, {\tilde E_c}$, is then proportional to the product
of $W$ and ${\tilde W}$ evaluated at the horizon.
In section 3 we turn to the discussion of the FRW cosmology on the
brane, following \cite{Savonije:2001nd}.  
Using 
\cite{Balasubramanian:1999re,Emparan:1999pm,Kraus:1999it,Kraus:1999di,
Batrachenko:2004fd} 
we determine the equations
describing the motion of the dual brane in the black hole background.
We briefly discuss the cosmology on the brane
in the background of 
one of the black hole
solutions of the STU model, and we comment on
the Bekenstein bound.
Then we show that the FRW equations have a form
that is similar to the equations for the entropy and
the Casimir energy on the brane, and that
both sets coincide 
when the brane crosses the event horizon.
In section 4 
we conclude with a few comments on the inclusion of a Gauss-Bonnet term
in the analysis.
See
also \cite{Nojiri:2001ae,Lidsey:2002zw,Nojiri:2002hz,Cai:2002bn,
Gregory:2003px} 
for related discussions.

\section{
Non-extremal
electrically charged static black holes in \\
$N=2$ gauged supergravity theories }

In the following, we consider $N=2$ gauged supergravity 
theories in five dimensions
obtained by gauging 
the $U(1)$ subgroup of the $SU(2)$-automorphism group of
the $N=2$ supersymmetry algebra \cite{Gunaydin:1984ak} .  The gauging is 
with respect to a linear combination 
proportional to $h_A \, A_M^A$ 
of $U(1)$ gauge fields
(with constant $h_A$), 
and the coupling constant $\mathfrak{g}$ is identified with the
inverse of the curvature radius of $AdS_5$, i.e. $\mathfrak{g} = L^{-1}$.

The relevant part of the five-dimensional action reads 
\cite{Gunaydin:1984ak}
\begin{equation}
S = \frac{1}{16 \pi \, G_5} \int d^5x \sqrt{-g} \left( R 
-  g_{ij} \, \partial_M \varphi^i \partial^M \varphi^j 
- \frac12 G_{AB} \,F^A_{MN} F^{B \,MN } - 
V_{\rm pot}
\right) 
\;.
\label{bulkaction5}
\end{equation}
We denote the spacetime metric by $g_{MN}$.
The real scalar fields $X^A$ satisfy the constraint
\begin{equation}
\frac16 \, C_{ABC} \, X^A \, X^B \, X^C =1 \;.
\label{vsgconstraint}
\end{equation}
The metric $G_{AB}$ is given by
\begin{equation}
G_{AB} = - \frac12 \, C_{ABC} \, X^C + \frac92 \, X_A \, X_B \;,
\end{equation}
where
\begin{equation}
X_A = \frac16 \, C_{ABC} \, X^B \, X^C \;.
\end{equation}
Observe that $X^A \, X_A =1$ in view of \eqref{vsgconstraint}.
In addition,
\begin{equation}
X_A \, \partial_i X^A =0 \;,
\label{vsgxdx}
\end{equation}
where $X^A = X^A (\varphi^i)$ and 
$\partial_i X^A (\varphi) = \partial X^A/ \partial \varphi^i$.  Here
the $\varphi^i$ denote the physical scalar fields with target-space
metric
\begin{equation}
g_{ij} = G_{AB} \, \partial_i X^A \, \partial_j X^B \;.
\end{equation}
A useful relation is 
\begin{equation}
G_{AB} \, \partial_i X^B = - 
\frac32 \, \partial_i X_A \;.
\label{gderxx}
\end{equation}
The potential $V_{\rm pot}$, which is expressed in terms of 
the superpotential
\begin{equation}
W =  h_A \, X^A \;,
\label{supo}
\end{equation}
reads
\begin{equation}
V_{\rm pot} = \mathfrak{g}^2 
\left( g^{ij} \, \partial_i W \,\partial_j W - 
\frac43 \, W^2 \right)
= 
\mathfrak{g}^2 \left( h_A \, G^{AB} \, h_B - 2 \, W^2 \right)
\;,
\label{potent}
\end{equation}
where in the second step we used
\begin{equation}
g^{ij}\, \partial_i X^A \,\partial_j X^B = G^{AB} - \frac23 \, X^A \, X^B \;.
\label{gijGab}
\end{equation}

\subsection{First-order flow equations}

The equations of motion derived from \eqref{bulkaction5} allow
for various classes of solutions that have a description in terms of 
first-order flow equations. In the ungauged case ($ \mathfrak{g} =0$)
one such class consists of electrically 
charged static extremal black hole solutions
with line element \cite{Ferrara:1997tw,Sabra:1997yd,Larsen:2006xm}
\begin{equation}
ds^2_5 = - {\rm e}^{-4 U} \,  dt^2 + {\rm e}^{2 U} \,  dr^2
+  {\rm e}^{2 U} \, r^2 \, d\Omega_3^2 \;.
\label{linebhung}
\end{equation}
The metric factor ${\rm e}^{2U}$ and the scalar fields $\varphi^i$
supporting the spherically symmetric black hole solution 
only depend on the radial coordinate $r$.  They satisfy the first-order
flow equations 
\begin{eqnarray}
\frac{d {\rm e}^{2U}}{d \xi} &=& \frac13 \, Z \;, \nonumber\\
\frac{d \varphi^i}{d \xi} &=& - \frac12 \, {\rm e}^{-2U} \, g^{ij} \,
\partial_j Z \;,
\label{floweqZ}
\end{eqnarray}
where $\xi$ denotes the variable $\xi = 1/r^2$ and where
$Z= q_A \, X^A$.  These flow equations can be combined into
\begin{equation}
X_A' + 2 U' \, X_A = - \frac23 \, \ee^{-2U} \, \frac{q_A}{r^3} \;,
\label{flowbhung}
\end{equation}
where $' = d/dr$.  Indeed, contracting \eqref{flowbhung} with $X^A$
results in the flow equation for ${\rm e}^{2U}$, while contracting with
$\partial_j X^A$ yields the flow equation for $\varphi^i$ in view
of the very special geometry relations \eqref{vsgxdx} and 
\eqref{gderxx}.

The flow equations (\ref{floweqZ}) are solved in terms of harmonic
functions $H_A$,
\begin{eqnarray}
{\rm e}^{2U} &=& \frac13 \, H_A \, X^A \;, \nonumber\\
{\rm e}^{2U} \, X_A &=& \frac13 \, H_A \;, 
\label{solflowZ}
\end{eqnarray}
where $H_A = c_A + q_A/r^2$, and where the $c_A$ denote arbitrary 
integration constants.

In the gauged case ($ \mathfrak{g} \neq 0$) a well-known 
class of solutions admitting
a description in terms of first-order flow equations are flat domain wall
solutions with line element \cite{Girardello:1998pd,Freedman:1999gp,
Behrndt:1999ay,Behrndt:1999kz,
Skenderis:1999mm,Kallosh:2000tj,Behrndt:2000zh,Behrndt:2000tr,Gubser:2000nd,
Ceresole:2001wi}
\begin{equation}
ds^2_5 = {\rm e}^{2 A} \, \eta_{\mu \nu} \, dx^{\mu} dx^{\nu} 
+ d \rho^2 \;.
\end{equation}
The metric factor ${\rm e}^{2A}$ and the scalar fields $\varphi^i$
supporting the domain wall solution 
only depend on the radial coordinate $\rho$. 
They satisfy the first-order
flow equations 
\begin{eqnarray}
\frac{ d A}{d \rho} &=& \frac13 \, \mathfrak{g} 
\, W \;, \nonumber\\
\frac{d \varphi^i}{d \rho} &=& - \mathfrak{g} \, g^{ij} \, \partial_j W \;.
\label{flowdw}
\end{eqnarray}
Changing the radial variable from $\rho$ to $r$ such that
$d \rho/ dr = ( \mathfrak{g} \, r \, {\rm e}^{2 U (r)})^{-1}$, with 
${\rm e}^{A (\rho)} = \mathfrak{g} \, r \, {\rm e}^{U (r)} $, yields the 
line element in the form
\begin{equation}
ds^2_5 = {\rm e}^{- 4 U} \, f \, \eta_{\mu \nu} \, dx^{\mu} dx^{\nu} 
+ {\rm e}^{2 U} \, f^{-1} \, d r^2 \;,
\label{linedwgau}
\end{equation}
where
\begin{equation}
f = \mathfrak{g}^2 \, r^2\, {\rm e}^{6 U} \;.
\end{equation}
The flow equations \eqref{flowdw} now take the form
\begin{eqnarray}
U' \, r &=& \frac13 \, {\rm e}^{-2 U} \, W - 1 \;,\nonumber\\
X'^A &=& \frac{\ee^{-2U}}{r} \left( \frac23 \, W \, X^A - G^{AB} h_B
\right)  \:,
\label{floweqw}
\end{eqnarray}
where $' = d/dr$.  Here we have displayed the flow equation for the $X^A$.
The flow equation for the $\varphi^i$, 
\begin{equation}
\varphi^i{}' = - \frac{{\rm e}^{-2U}}{r} 
\, g^{ij} \, \partial_j W \;,
\end{equation}
 follows from the flow equation for $X^A$ by contracting it with
$G_{AB} \, \partial_j X^B$.

The electrically charged black hole solutions of ungauged supergravity
described above satisfy first-order flow equations based on $Z = q_A \, X^A$,
whereas the flat domain wall solutions of gauged supergravity just described
satisfy first-order flow equations based on $W = h_A \, X^A$.  One
may ask whether there exist charged solutions to gauged supergravity
satisfying both sets of first-order flow equations \eqref{flowbhung} and
\eqref{floweqw} (common features of black hole and domain wall solutions
were recently discussed in \cite{Ceresole:2007wx}).
That there exist charged extremal 
solutions in gauged supergravity
based on first-order flow equations was demonstrated in \cite{Elvang:2007ba},
where various examples with one real scalar field and one abelian
gauge field were discussed.  
Non-extremal electrically
charged static black hole solutions were constructed in \cite{Behrndt:1998jd} 
by solving the equations of motion.
Here we show that these solutions 
have a first-order flow description based on the two sets
\eqref{flowbhung} and \eqref{floweqw}, by rewriting the five-dimensional
action \eqref{bulkaction5} in terms of these equations.
Then, the 
compatibility of the flow equations \eqref{flowbhung} and \eqref{floweqw}
requires identifying the integration constants $c_A$ appearing in the solution
\eqref{solflowZ} 
 with $h_A$, so that now
\begin{equation}
H_A = h_A + \frac{q_A}{r^2} \;.
\label{harmf}
\end{equation}

Following \cite{Behrndt:1998jd} we consider non-extremal electrically charged
static black hole solutions with line element
\begin{equation}
ds^2_5 = - {\rm e}^{-4 U} \, f \, dt^2 + {\rm e}^{2 U} \, f^{-1} \, dr^2
+  {\rm e}^{2 U} \, r^2 \, d \Sigma_k^2 \;\;\;,\;\;\;
f = k - \frac{\mu}{r^2} + \mathfrak{g}^2 \, r^2 \, {\rm e}^{6 U} \;,
\label{bhline}
\end{equation}
where $U=U(r), f=f(r)$.  
Here $d \Sigma_k^2$ denotes the line element of 
a three-dimensional space of constant curvature with metric 
$\eta_{\alpha \beta}$, either flat space ($k=0$), hyperbolic space ($k=-1$)
or a unit three-sphere $S^3$ ($k=1$). 
The presence of a non-vanishing parameter $\mu$
is necessary in order for the solutions to have a horizon.
Observe that the line elements \eqref{linebhung} and
\eqref{linedwgau} are special cases of \eqref{bhline}.
In the following, we will always consider the case $k=1$, but we keep
$k$ in the formulae as a book-keeping device.
The scalar fields and the gauge fields supporting the solutions are
taken to be functions of $r$, only.
Inserting the line element \eqref{bhline} into the action 
\eqref{bulkaction5} yields
\begin{equation}
16 \pi \, G_5 \, S = S_0 + S_2 + S_{\rm td} \;,
\end{equation}
where $S_0$ and $S_2$ comprise the contributions to order $\mathfrak{g}^0$ 
and $\mathfrak{g}^2$, respectively,
and where $S_{\rm td}$ contains total derivative terms.  $S_0$ and $S_2$ read
\begin{eqnarray}
S_0 &=& \int d^5x \sqrt{\eta} \left[ 3 \,\mu \, \frac{\ee^{-2U}}{r^3} \, q_A G^{AB} \left(2 X_B - \frac13 \,
\ee^{-2U} \, H_B \right)
\right. \nonumber\\
&& 
 - \frac94 \, \left( k - \frac{\mu}{r^2} \right) \, r^3 \, 
\left( X_A' + 2 U' \, X_A + \frac23 \, \ee^{-2U} \, \frac{q_A}{r^3} \right) G^{AB}
\left( X_B' + 2 U' \, X_B + \frac23 \, \ee^{-2U} \, \frac{q_B}{r^3} \right) 
\nonumber\\
&& \left. + r^3 \, \ee^{4 U} \left( F^A_{tr} - \ee^{-4U} \, \frac{G^{AC}  Q_C}{r^3} \right) G_{AB}
\left( F^B_{tr} - \ee^{-4U} \, \frac{G^{BD}  Q_D}{r^3}
\right)
\right] \;, \nonumber\\
S_2 &=& \mathfrak{g}^2 \int d^5x \sqrt{\eta} \left[ \right. 
\frac43 \, r^3 \, \ee^{2 U} \left(W - 3  \, \ee^{2U} \, \left(U' r + 1 
\right) \right)^2
\nonumber\\
&& 
- r^5 \, \ee^{6 U} \left( X'^A - \frac{\ee^{-2U}}{r} \left( \frac23 \, W
\, X^A - G^{AC} h_C \right) \right)
G_{AB} \nonumber\\
&& \left. \qquad \qquad \qquad 
\left( X'^B - \frac{\ee^{-2U}}{r}
\left( \frac23 \, W \, X^B - G^{BD} h_D \right) \right)
\right] \;,
\end{eqnarray}
where $' = d/dr$.
In the expression for $S_0$,
the physical electric charges $Q_A$ 
are related to the $q_A$ by
\begin{equation}
Q_A G^{AB} Q_B = k \, q_A G^{AB} q_B + \mu \, q_A G^{AB} h_B \;,
\label{Qq}
\end{equation}
and $H_A$ is given by \eqref{harmf}.
$S_{\rm td}$ reads
\begin{eqnarray}
S_{\rm td} &=&  \int d^5x \sqrt{\eta} \, \left[ 2 \, Q_A \, F^A_{tr} 
+ \left(6 \, \mu \, U -2
(k - \frac{\mu}{r^2}) \, ( r^3 \, U'
+ \ee^{- 2U} \, q_A X^A  )
  \right)' \right. \nonumber\\
&& \left. + \mathfrak{g}^2  \left(  -8 r^5 \, \ee^{6U} \, U' 
- 8 r^4 \, \ee^{6U} + 2 r^4 \, \ee^{4U} \, W \right)'
\right] \;.
\label{stder}
\end{eqnarray}
Observe that $S_0$ 
and $S_2$ are given in terms of squares of 
first-order differential equations,
with
the exception of the first line of $S_0$, which is proportional to the 
parameter $\mu$.  
Under variation with respect
to $U$ or with respect to $X^A$, the contribution from 
the first line of $S_0$ vanishes provided that 
\begin{equation}
X_A = \frac13 \, \ee^{-2 U} \, H_A \;.
\label{xa}
\end{equation}
Thus, extremizing $S_0$ and $S_2$ with respect to 
the fields $U$, $X^A$ and $F^A_{tr}$ yields
the relation \eqref{xa},  
the first-order flow equations
\eqref{flowbhung} and \eqref{floweqw}
as well as 
\begin{equation}
F^A_{tr} = \ee^{-4U} \, \frac{G^{AC}  Q_C}{r^3} \;.
\label{fieldstr}
\end{equation}
The flow equations 
\eqref{flowbhung} and \eqref{floweqw}
are solved by
\eqref{solflowZ} with $c_A = h_A$, 
as discussed above, and the solution agrees with
\eqref{xa}.  

We take $h_A$ and $q_A$ in \eqref{harmf} to be positive
to ensure that $H_A >0$.  We also take $X^A > 0$, so that
$\ee^{2U} >0$ along the flow.
We impose the normalization $\ee^{2 U} = 1$ at $r = \infty$.  
The asymptotic value  of $X_A$
is then $ \frac13 \, h_A$.  Denoting 
the asymptotic value of the $X^A$ by $h^A$, we have $ \frac13 h^A \, h_A =1$
in view of \eqref{vsgconstraint}.  We introduce the `dual'
superpotential
${\tilde W}$ as 
\begin{equation}
{\tilde W} = h^A \, X_A \;,
\label{dualsupo}
\end{equation} 
for later convenience.

Summarizing, the line element \eqref{bhline} together with
\eqref{solflowZ}, \eqref{harmf}, \eqref{fieldstr} 
and \eqref{Qq} describe non-extremal electrically charged static black hole
solutions to first-order flow equations.   
It can be checked that they solve all the equations of motion.

On the solution, the total derivative terms \eqref{stder} can be written as
\begin{equation}
S_{\rm td} = \int d^5x \sqrt{\eta} \left[6 \, \mu \, U' - 2 \left(
r^3 \, f \, U' + r^2 (f -k) \right)' \right] \;,
\end{equation}
while the first line of $S_0$ yields
\begin{equation}
- \int d^5x \sqrt{\eta} \, 6 \, \mu \, U' \;.
\end{equation}
Thus, on the solution, the action \eqref{bulkaction5} evaluates to
\begin{equation}
16 \pi \, G_5 \, S = -2 \int d^5x \sqrt{\eta} \left(
r^3 \, f \, U' + r^2 (f -k) \right)' \;,
\label{valuebulksol}
\end{equation}
in agreement with \cite{Batrachenko:2004fd}.

The electric field $F_{tr}^A$ is determined in terms of the potential 
${\phi}^A (r)$, i.e. $F_{tr}^A = - 
\partial_r {\phi}^A (r)$.  
In the following, we compute the contraction $Q_A \, {\phi}^A$.
To this end, we differentiate the first equation of \eqref{floweqZ}
and obtain
\begin{equation}
U'' =  - \frac13 \, \ee^{-2U} \, \left( - 2 U' \,\frac{Z}{r^3} - 3 \frac{Z}{r^4} + \frac{q_A X'^A}{r^3} \right) \;.
\label{udd}
\end{equation}
Using \eqref{flowbhung} we compute
\begin{equation}
X'^A = - \frac32 \, G^{AB} \, X'_B = 2 U' \, X^A + \ee^{-2U} \, \frac{G^{AB} q_B}{r^3} \;.
\label{dxd}
\end{equation}
Then, using \eqref{udd} and \eqref{dxd} gives
\begin{equation}
U'' + \frac{3}{r} \, U' = - \frac13 \, \ee^{-4U} \, \frac{q_A G^{AB} q_B}{r^6} \;.
\label{uddup}
\end{equation}
With the help of \eqref{floweqZ}, \eqref{Qq} and
\eqref{uddup} we obtain
\begin{equation}
- 3 \left[ r^3 \left(k - \frac{\mu}{r^2}\right) U' \right]' = \ee^{-4U} \, \frac{Q_A G^{AB} Q_B}{r^3} \;,
\end{equation}
which equals $Q_A \, F^A_{tr}$, as can be seen from \eqref{fieldstr}.
Hence we establish that
\begin{equation}
Q_A \, {\phi}^A =  - \left(k - \frac{\mu}{r^2}\right) \, \ee^{-2U} \, 
q_A X^A + k \, q_A h^A \;,
\label{qpourn}
\end{equation}
where we used the first equation of \eqref{floweqZ} once more.
We chose the integration constant in such a way that $Q_A \, 
{\phi}^A$ vanishes 
at spatial infinity, as in \cite{Gauntlett:1998fz}.
In the context of the AdS/CFT correspondence
\cite{Maldacena:1997re,Gubser:1998bc,Witten:1998qj}
this means that the potentials ${\phi}^A$ associated with
the $U(1)$ charges $Q_A$
approach the boundary at a vev rate 
\cite{Bianchi:2001kw,Elvang:2007ba}.   A different choice of the integration
constant can be incorporated into the Cardy-Verlinde-type formula given below
by an appropriate shift.

The quantity appearing in the first law of black hole mechanics, 
$d M = T_H \,d {\cal S} + \phi^A \,d Q_A$, 
is not \eqref{qpourn} but a rescaled
one given by
\cite{Chamblin:1999tk}
\begin{equation}
Q_A \, \phi^A =  
\frac{2}{3 w_5} \left(  - \left(k - \frac{\mu}{r^2}\right) \, \ee^{-2U} \, 
q_A X^A +  k \, q_A h^A  \right)\;,
\label{qphi}
\end{equation}
where
\begin{equation}
w_5 = \frac{16 \pi}{3} \frac{ G_5}{ {vol}} \;.
\label{defw5}
\end{equation}
Here, ${vol} = \int d^3x \sqrt{\eta}$ denotes the volume of the
three-dimensional space of constant curvature with line element $d \Sigma_k^2$.
For $k=1$ this space
is a unit three-sphere $S^3$ with volume 
$ {vol} = {vol} (S^3)$.  As already stated,
we will focus on the case $k=1$ 
but we will nevertheless keep $k$ in the formulae as a book-keeping device.

Next, we compute the coefficient of 
the $1/r^{2}$-term
in the metric factor 
$- {\rm e}^{-4 U} \, f $ of the line element \eqref{bhline}.
We denote this coefficient by $w_5 \, M$.
Expanding $\ee^{2 U} = 1 + \kappa/r^2 + \dots$ 
as well as $X^A = h^A + \beta^A/r^2 + \dots$ and using the first equation of 
\eqref{solflowZ}
results in
$\kappa = \frac13 \left(h^A \, q_A + h_A \, \beta^A \right) $.  
On the other hand, inserting
the expansion of $X^A$ into \eqref{vsgconstraint} yields $h_A \, \beta^A = 0$.  It follows that the coefficient $w_5 \, M$
is given by
\begin{equation}
w_5 \, M = \mu + 2 k \, \kappa = \mu + \frac23 \, k \, h^A \, q_A \;.
\label{adm}
\end{equation}
$M$ denotes the mass of the black hole relative to the asymptotic $AdS_5$ 
spacetime.  Pure global $AdS_5$ ($k=1$) has a mass given by $\frac14 \,
w_5^{-1} \, \mathfrak{g}^{-2}$ \cite{Balasubramanian:1999re}, so
that the total energy reads \cite{Balasubramanian:1999re,Batrachenko:2004fd}
\begin{equation}
\frac{1}{w_5} \left( \mu + \frac23 \, h^A \, q_A + \frac{1}{4 
\, \mathfrak{g}^{2}} \right) \;.
\end{equation}
The contribution proportional to $1/\mathfrak{g}^{2}$ will, however, not play
any role in the Cardy-Verlinde formula and in the matching of the
FRW equations with thermodynamic equations on the brane.

Next, let us discuss a few black hole solutions explicitly.  We
specialize \eqref{vsgconstraint} to the case of the STU model
with $X^1 X^2 X^3 =1$.  The associated metric $G_{AB}$ is diagonal.
The solution to the equations 
\eqref{solflowZ} reads \cite{Sabra:1997yd}
\begin{equation}
{\rm e}^{6 U} = H_1 H_2 H_3 \;\;\;,\;\;\; X^1 = 
\frac{\left(H_2 H_3 \right)^{1/3}}{H_1^{2/3}} \;\;\;,\;\;\;
 X^2 = 
\frac{\left(H_1 H_3 \right)^{1/3}}{H_2^{2/3}} \;\;\;,\;\;\;
 X^3 = 
\frac{\left(H_1 H_2 \right)^{1/3}}{H_3^{2/3}} \;.
\end{equation}
The Maxwell case 
is obtained by 
setting $H_1 =H_2 = H_3 = 1 + q/r^2$.  Then
$X^1 = X^2 = X^3 = 1$ and $\ee^{2U} = H_1$, so that $W=3 \,, {\tilde W}=1$, and
\eqref{adm} and 
\eqref{qphi} become (with $k=1$)
\begin{equation}
w_5 \, M = \mu + 2 q \;\;\;,\;\;\;
Q_A \, \phi^A  = \frac{2}{w_5} \, \frac{q (\mu + q)}{q + r^2}  \;.
\label{phimax}
\end{equation}
On the other hand, setting $H_1 =H_2 = 1 + q/r^2$ and $H_3 =1$
yields the $p=\frac23$ example discussed in \cite{Elvang:2007ba} 
with
${\rm e}^{6 U} = H^2_1$ and $X^1 = X^2 = H^{-1/3}_1 \,,\, X^3 = H^{2/3}_1$.
Then $W = 2 H^{-1/3}_1 + H^{2/3}_1$ and ${\tilde W} = \frac13 \left(2 H^{1/3}_1
+H^{-2/3}_1 \right)$,
 and \eqref{adm} and 
\eqref{qphi} become (with $k=1$)
\begin{equation}
w_5 \, M = \mu + \frac43 \, q \;\;\;,\;\;\;
Q_A \, \phi^A  = \frac{4}{3 w_5} \, \frac{q (\mu + q)}{q + r^2}  \;.
\label{solfreedp23}
\end{equation}
The existence of a horizon 
shielding the singularity at $r=0$ 
requires taking $ \mu > \mathfrak{g}^2 \, q^2$,
and there is no inner horizon \cite{Gubser:2004xx,Elvang:2007ba}.

Finally, setting $H_1 = 1 + q/r^2$ and $H_2 = H_3 =1$
yields the $p=\frac13$ example discussed in \cite{Elvang:2007ba} 
with
${\rm e}^{6 U} = H_1$ and $X^1 = H^{-2/3}_1 \,,\,
X^2 = X^3 = H^{1/3}_1 $.
Then $W = H^{-2/3}_1 + 2 H^{1/3}_1$ and ${\tilde W} = \frac13 \left(H^{2/3}_1
+2 H^{-1/3}_1 \right)$,
 and \eqref{adm} and 
\eqref{qphi} become (with $k=1$)
\begin{equation}
w_5 \, M = \mu + \frac23 \, q \;\;\;,\;\;\;
Q_A \, \phi^A  = \frac{2}{3 w_5} \, \frac{q (\mu + q)}{q + r^2}  \;.
\label{free2}
\end{equation}
The solution has a single horizon shielding the singularity at $r=0$
whenever $\mu > 0$ \cite{Gubser:2004xx,Elvang:2007ba}.

\subsection{A Cardy-Verlinde-type formula for charged black holes}

Now we will show that for any model \eqref{vsgconstraint} 
the entropy of a charged black hole
\eqref{bhline} 
can be written as a Cardy-Verlinde-type
formula \cite{Verlinde:2000wg}.  This has already been discussed
in \cite{Cai:2001jc,Biswas:2001sh,Klemm:2001pn,Youm:2001qr,Gregory:2002am}
for various black holes in the context of the STU model.

In the coordinates \eqref{bhline}, the exterior
horizon $r_h$ of the black hole 
is located at the largest real positive root of 
(we assume ${\rm e}^{- 4 U(r_h)} \neq 0$)
\begin{equation}
f(r_h) =0 \;.
\label{hor}
\end{equation}
Its Hawking temperature $T_H$ is given by
\cite{Batrachenko:2004fd}
\begin{equation}
T_H = \frac{1}{4 \pi}  \, f'(r_h)  \, \ee^{-3 U(r_h)}  \;.
\label{thawk}
\end{equation}
In the following, we consider a black hole with a spherical horizon
($k=1$).
The Bekenstein-Hawking entropy of the black hole is given by a quarter of the 
area of the event horizon, a three-sphere with radius $a_h = r_h \, \ee^{U(r_h)}$, 
\begin{equation}
{\cal S} = \frac{{vol} (S^3) \, a^3_h }{4 \, G_5} = 
\frac{4 \pi}{3 w_5} \, a^3_h \;,
\label{entro}
\end{equation}
with $w_5$ given by \eqref{defw5}.
Hence
\begin{equation}
T_H \, {\cal S} = \frac{1}{3 w_5} \, f'(r_h) \, r^3_h \;,
\end{equation}
which we now compute.  Using the first equation of \eqref{floweqZ}
we obtain
\begin{equation}
f'(r) \, r^3  = 2 \mu + 2 \, \mathfrak{g}^{2} 
\,r^4 \,\ee^{6U} \left(1 - \ee^{-2U} \, \frac{q_A X^A}{r^2} \right) \;.
\end{equation}
At the horizon, it follows from \eqref{hor} that
\begin{equation} 
\mu - k \, r^2_h = \mathfrak{g}^{2} \, r^4_h \, \ee^{6U (r_h)} \;,
\label{horrel}
\end{equation}
and hence
\begin{equation}
T_H \, {\cal S} = \frac{2}{3 w_5} \, \left[2 \mu - k \, 
r^2_h + \left(k - \frac{\mu}{r^2_h} \right) \ee^{-2U(r_h)}
\, q_A X^A_h \right]\;.
\label{tshor}
\end{equation}

Next, let us consider the Smarr-type combination
\begin{equation}
\frac43 M - T_H \, {\cal S} - Q_A \, \phi^A_h 
\;,
\label{casi}
\end{equation}
with $Q_A \, \phi^A_h$ given by \eqref{qphi}
and evaluated at the horizon. 
This combination is the gravitational counterpart
of the Casimir energy $E_c/3$ on the brane.  
$E_c$ is defined as the violation of
the thermodynamic Euler relation {\cite{Verlinde:2000wg}, as we will
briefly review in the next section.
The combination \eqref{casi} can also be motivated by exhibiting 
its relation to the Smarr formula, as follows. 
In the absence of charges, the area $A$
of the event horizon is determined in terms of the mass parameter $\mu$
and $\mathfrak{g}$ using \eqref{horrel}.  
We can view this as a relation
$\mu = \mu (A, \mathfrak{g})$. 
Under the 
simultaneous 
rescaling $r_h \rightarrow
\lambda \, r_h$ and 
$\mathfrak{g} \rightarrow
\lambda^{-1} \, \mathfrak{g}$ 
we have $A \rightarrow \lambda^3 \, A$ as well as
$\lambda^2 \, \mu =
\mu (\lambda^3 \, A, \lambda^{-1} \,\mathfrak{g})$.
Differentiating with
respect to $\lambda$, setting $\lambda = 1$ and multiplying with $w_5$
results in
\begin{equation}
2 M = 3 T_H \, {\cal S} - \mathfrak{g} \, \frac{\partial  M}{\partial
\mathfrak{g}} \;,
\label{smarrmod}
\end{equation}
where we used the first law of thermodynamics, $d M = T_H \, d {\cal S}$.
Using \eqref{horrel}, we compute 
\begin{equation}
\mathfrak{g} \, \frac{\partial  M}{\partial \mathfrak{g}} = 2 M - 
\frac{2 k}{w_5} \, r^2_h \;.
\label{gmg}
\end{equation}
Inserting \eqref{gmg} into \eqref{smarrmod} we obtain
\begin{equation}
\frac43  M - T_H \, {\cal S} =
\frac{2 k}{3 w_5} \, r^2_h \;.
\label{smarrmodschwarz}
\end{equation}
This is the result for 
the Smarr-type combination \eqref{casi} for uncharged black holes.
In the ungauged case ($\mathfrak{g} =0$) we have
 $k \, r^2_h = \mu$ and \eqref{smarrmodschwarz}
yields the Smarr formula $\frac23  M = T_H \, {\cal S}$ \cite{Smarr:1972kt}.

In analogy to \cite{Verlinde:2000wg},
we will denote the Smarr-type combination \eqref{casi} by  
${\tilde E}_c/3$.
Using \eqref{adm}, \eqref{tshor} and \eqref{qphi} we obtain
\begin{equation}
{\tilde E}_c = \frac{2\,k}{w_5} \,  
\left(r^2_h + \frac13 \, h^A q_A\right) \;.
\label{casim}
\end{equation}
On the other hand, contracting \eqref{xa} with $h^A$ gives
\begin{equation}
r^2 \ee^{2U} \, h^A X_A = r^2 + \frac13 h^A q_A  \;,
\label{ruhq}
\end{equation}
and hence
\begin{equation}
{\tilde E}_c = \frac{2\, k}{w_5} \, {\tilde W}_h \, a^2_h \;,
\label{tildeEcw}
\end{equation}
where ${\tilde W}_h$ denotes the `dual' superpotential 
\eqref{dualsupo} evaluated
at the horizon.  The quantity ${\tilde E}_c$ is thus non-vanishing for
a horizon of spherical topology.
Observe that $a^2_h \propto {\cal S}^{2/3}$
(cf. \eqref{scaleeec}). 
In the ungauged case ($\mathfrak{g} =0$) we have
 $k \, r^2_h = \mu$ as well as ${\tilde E}_c = 2 M - Q_A \, \phi^A_h$,
and \eqref{casi} yields the Smarr formula $\frac23  M = T_H \, {\cal S}
+ \frac23 Q_A \, \phi^A_h$ \cite{Smarr:1972kt,Gauntlett:1998fz}.

In the gauged case  ($\mathfrak{g} \neq 0$) the combination
$2 M - Q_A \, \phi^A_h - {\tilde E}_c$ is no longer vanishing.
We find
\begin{equation}
2 M - Q_A \, \phi^A_h - {\tilde E}_c = \frac{2}{3 w_5} \,(\mu - k \, r^2_h)  
\left(3 - \ee^{-2U (r_h)} \frac{q_A X^A_h }{r^2_h} \right) \;,
\end{equation}
where we used \eqref{adm}, \eqref{qphi} and \eqref{casim}. With the help of 
the first equation of \eqref{solflowZ}
and \eqref{supo} we have
\begin{equation}
\frac{q_A X^A }{r^2} = 3 \ee^{2U} - W \;,
\label{relqUW}
\end{equation}
so that 
\begin{equation}
2 M - Q_A \, \phi^A_h - {\tilde E}_c = \frac{2}{3 w_5} \,(\mu - k \, r^2_h)  \,
\ee^{-2U (r_h)} W_h \;,
\end{equation}
where $W_h$ denotes \eqref{supo} evaluated at the horizon.
Using \eqref{horrel} this can be written as
\begin{equation}
2 M - Q_A \, \phi^A_h - {\tilde E}_c = \frac{2}{3 w_5} \, \mathfrak{g}^{2} \,
W_h  \,a^4_h \;,
\label{2mEcw}
\end{equation}
which is positive.
Observe that $a^4_h \propto {\cal S}^{4/3}$
(cf. \eqref{scaleeec}). We denote this combination 
by $2 \, {\tilde E}_e$, and we note that $2 M = 2 \, {\tilde E}_e
+ {\tilde E}_c +  Q_A \, \phi^A_h$.

It follows that we can express the square of the entropy \eqref{entro}
as
\begin{equation}
k \, \mathfrak{g}^{2} \,
W_h \, {\tilde W}_h \, {\cal S}^2 = 
\frac{8 \pi^2}{3 }  
\, {\tilde E}_e \, 
{\tilde E}_c
= 
\frac{4 \pi^2}{3 }  
\, 
{\tilde E}_c \left(
2 M - Q_A \, \phi^A_h - {\tilde E}_c \right) \;.
\label{cvbh}
\end{equation}
This is a Cardy-Verlinde-type formula for charged AdS 
black holes, here expressed in terms of gravitational
quantities. It makes use of both $W$ and ${\tilde W}$ evaluated at the
horizon.  
Using \eqref{gijGab} we note the relation
\begin{equation}
W \, {\tilde W} = 3 \left( 1 + \frac12 \,
g^{ij} \, \partial_i W \, \partial_j {\tilde W}
\right) \;,
\label{wtilw3}
\end{equation}
which shows that in general $W_h \, {\tilde W}_h \neq 3$.

Summarizing, we find that three combinations are naturally expressed in terms
of the superpotential quantities $W$ and ${\tilde W}$, namely 
\eqref{tildeEcw}, \eqref{2mEcw} and \eqref{cvbh}.

\section{FRW cosmology on the brane}

A radiation dominated FRW universe can be described in terms of a 
co-dimension-one 
brane of fixed tension moving in the background of a five-dimensional
AdS-Schwarzschild black hole \cite{Gubser:1999vj,
Verlinde:2000wg,Savonije:2001nd}.
The radiation is represented by a CFT on the brane, and 
the induced metric on the brane takes the form of a standard FRW metric.
The FRW equations have a structure that is similar to the 
equations for the entropy and the Casimir energy
on the brane, and both sets of equations coincide 
when the brane crosses the horizon of the black hole
\cite{Verlinde:2000wg,Savonije:2001nd}.  
This also holds for the charged static 
AdS black holes 
discussed in the previous section, as we will show below.

\subsection{Brane motion in the background
of an AdS black hole}

Following \cite{Gubser:1999vj,deHaro:2000wj,Savonije:2001nd} 
we regard the brane
as the boundary of the AdS geometry.
The bulk-spacetime action \eqref{bulkaction5}
needs to be supplemented by a boundary term,
the so-called Gibbons-Hawking term \cite{Gibbons:1976ue},
and by counterterms
\cite{Balasubramanian:1999re,Emparan:1999pm,Kraus:1999di,Batrachenko:2004fd},
\begin{equation}
- \frac{1 }{8 \pi \, G_5} \int_{\Sigma} d^4 x \sqrt{- \gamma}
\left(K + \mathfrak{g} \, W
+ \frac{1}{4 \, \mathfrak{g}} \, {\cal R}
 \right) \;,
\label{combinedaction5}
\end{equation}
where $\Sigma$ 
denotes the brane, ${\cal R}$ denotes the Ricci scalar on the brane, and
$K$ is the Gibbons-Hawking term.
Observe that since we focus on the case $k=1$ for which 
the boundary geometry is $S^1 \times S^3$, the holographic
trace anomaly \cite{Henningson:1998gx} vanishes and no further counterterms
are required \cite{Emparan:1999pm,Burgess:1999vb}.
The counterterms ensure that when the brane is moved to infinity the 
on-shell action of the black hole, which is given by the sum of 
\eqref{valuebulksol} and 
\eqref{combinedaction5},
is finite.

The extrinsic curvature $K$ is given by
\begin{equation}
K = \gamma^{MN} \, K_{MN} \;\;\;,\;\;\; K_{MN} = \gamma_M{}^P \, 
\gamma_N{}^Q \, \nabla_{(P} \,
n_{Q)} \;.
\label{defKMN}
\end{equation}
Here the tensor 
$\gamma_{MN} = g_{MN} - n_M \, n_N$ 
denotes the projection of $g_{MN}$ onto $\Sigma$, 
so that the induced metric on $\Sigma$ is given by the tangential 
components of $\gamma_{MN}$.  We note that $\gamma^{MN} = g^{MN} - n^M \, n^N$
and $K  = \nabla_M  \, n^M $, where $n^M = g^{MN} \, n_N$.
The vector
$n = n^M \, \partial_M$ is the unit normal to $\Sigma$, 
i.e. $ n^M \, n_M = 1$.

We view the brane as a dynamical entity \cite{Gubser:1999vj,Savonije:2001nd} 
moving in a background of the form
\begin{equation}
ds^2_5 = g_{MN} \, dx^M \, dx^N = 
- A(a) \, dt^2 + B(a) \, da^2 + a^2 \, d\Sigma_k^2 \;.
\label{backgab}
\end{equation}
Note that the black hole metric \eqref{bhline}
is of this type with
\begin{equation}
a = r \, e^U 
\label{rarel}
\end{equation}
and
\begin{equation}
A = {\rm e}^{-4U} \, f \;\;\;,\;\;\;
B = \frac{1}{(1 + r \, U')^2 \; f}  \;\;\;,\;\;\; U' = 
\frac{dU}{dr} \;,
\label{fgh}
\end{equation}
where $A$ and $B$ are related by
\begin{equation}
A = \left( \frac{3}{W} \right)^2 \, \frac{1}{B} 
\label{relationAB}
\end{equation}
by virtue of the first equation of \eqref{floweqw}.

We take the induced metric $\gamma_{\mu \nu}$ on the brane
to have the form of a standard FRW metric
with cosmic scale factor $a (\tau)$ \cite{Savonije:2001nd},
\begin{equation}
ds^2_4 = 
- d \tau^2 + \gamma_{ij} \, dx^i dx^j =
- d \tau^2 + a^2 (\tau) \, d \Sigma^2_k \;.
\label{metricbrane}
\end{equation}
Comparing \eqref{metricbrane} with \eqref{backgab} shows that
the induced metric is obtained from the bulk metric
by requiring
\begin{equation}
- A \left(\frac{d t}{d \tau} \right)^2 + B \left( \frac{d a}{d \tau} \right)^2
= - 1 \:.
\label{relAB}
\end{equation}
This results in
\begin{equation}
\frac{d t}{d \tau} = A^{-1} \sqrt{A + A \, B \, {\dot a}^2} 
\;\;\;,
\;\;\; \dot a = \frac{ d a}{d \tau} 
\;.
\label{dtdta}
\end{equation}
To compute the associated vector $n^M$, we follow \cite{Kraus:1999it}
and introduce the velocity
vector $v^M$ ($v^M \, v_M =-1$),
\begin{equation}
v^M = \frac{d x^M}{d \tau} = ( \frac{dt}{d \tau}, \dot a , \vec{0}) 
\;.
\end{equation}
Using 
\begin{equation}
v^M \, n_M =0 \;,
\label{cond}
\end{equation}
we find that the unit normal vector $n^M$ is given by
\begin{equation}
n^M = \pm \frac{1}{\sqrt{A \,B}} ( B \, 
{\dot a}, \sqrt{ A + A\, B \,{\dot a }^2}, 
\vec{0} ) \;.
\label{normalvec}
\end{equation}
Below we will see that we have to take the minus sign for consistency.

Next, proceeding as in \cite{Chamblin:1999ya},
we vary the combined action \eqref{bulkaction5} 
and \eqref{combinedaction5}
with respect to the metric $g_{MN}$.
Setting the variation to zero
results in the equations
\cite{Brown:1992br,Balasubramanian:1999re,Chamblin:1999ya}
\begin{equation} 
K_{\mu \nu} - \left(K + \mathfrak{g} \, W \right) \,  \gamma_{\mu \nu} = 0 
\label{Keq}
\end{equation}
and
\begin{equation} 
{\cal R}_{\mu \nu} - \frac12 \, {\cal R} \, \gamma_{\mu \nu} = 8 \pi  \, G_4 \,
T_{\mu \nu}^{\rm matter} \;.
\label{inteq}
\end{equation}
Here we have split the brane equation of motion into two equations, where
the first one \eqref{Keq} is given in terms of the extrinsic curvature,
while the second one \eqref{inteq} is given in terms of the 
Ricci tensor 
on the brane $\Sigma$.  This splitting can be motivated by noting
that the five- and four-dimensional Newton's constants
are related by \cite{Gubser:1999vj}
\begin{equation}
G_5 = \frac12 \, G_4 \, L \;,
\label{g5g4}
\end{equation}
as can be seen from \eqref{combinedaction5}
(with $L = \mathfrak{g}^{-1}$).
Then, the terms in \eqref{inteq} stem from those terms in the
action \eqref{combinedaction5} that are proportional to $G_4^{-1}$
and hence intrinsically four-dimensional, 
whereas the terms in \eqref{Keq} come from terms in
\eqref{combinedaction5} that are multiplied by powers of $G_5$ and $L$
in such a way that these factors do not combine into
powers of $G_4$ only.   Thus,
$W$ only contributes to 
\eqref{Keq}, and there is no induced cosmological constant
on the brane (cf. \eqref{friedblack})
\cite{Savonije:2001nd}.
We note that in order
for the two equations \eqref{Keq} and \eqref{inteq}
to be consistent with one another,
we have to supplement the action \eqref{combinedaction5}
with terms describing the non-gravitational
degrees of freedom on the brane.  This results 
in the presence of 
an energy-momentum tensor
$T_{\mu \nu}^{\rm matter}$ in  \eqref{inteq} that is homogeneous
and isotropic, i.e.
$T^{\mu\,\rm matter}{}_{\nu} 
= {\rm diag}(-\rho_{\rm eff}, p_{\rm eff}, p_{\rm eff},
 p_{\rm eff}) $, and that is conserved,
\begin{equation}
\frac{d \rho_{\rm eff}}{d \tau} = - 3 H \, \left(\rho_{\rm eff} + 
p_{\rm eff} \right) \;,
\end{equation}
where  
\begin{equation}
H = \frac{\dot a}{a}
\end{equation}
denotes the Hubble parameter ($H$ should not be confused with the
$H_A$ given in \eqref{harmf}).

Tracing \eqref{Keq} yields
\begin{equation} 
K_{\mu \nu} =  - \frac{ \mathfrak{g}}{3} \, W \,\gamma_{\mu \nu}  \;,
\label{kmntracek}
\end{equation}
which we now evaluate.  
We first compute the $ij$-component of $K_{\mu \nu}$.  
Using the definition \eqref{defKMN} as well as \eqref{normalvec}
we obtain 
\begin{equation}
K_{ij} = \frac12 n^a \partial_a \gamma_{ij} = a^{-1} n^a \, \gamma_{ij} 
= \pm \frac{\sqrt{ A + A\, B {\dot a }^2}}{ \sqrt{ A\,B} \, a}  \, 
\gamma_{ij}\;,
\end{equation}
 and hence we get from  \eqref{kmntracek},
\begin{equation}
\pm \frac{\sqrt{ A + A \, B {\dot a }^2}}{ \sqrt{ A \, B} \, a} 
= - \frac{ \mathfrak{g}}{3} \, W
\;.
\label{dtdtauw}
\end{equation}
Since the right hand side is negative, we take the minus sign in
\eqref{normalvec}.
Then, squaring \eqref{dtdtauw} results in
\begin{equation}
H^2 = \left( \frac{ \mathfrak{g}}{3} \, W \right)^2 
- \frac{1}{B \, a^2} \;.
\label{hubble}
\end{equation}
This is the Friedmann equation describing the dynamics
of the scale factor $a (\tau)$.

Next, we compute the $\tau \tau$-component of \eqref{kmntracek}
using the method of 
\cite{Visser:1989kg,Chamblin:1999ya},
which we now review. We express $K_{\tau \tau}$
as
\begin{equation}
K_{\tau \tau} = K_{MN} \, v^M \, v^N = - n_N \, {\cal A}^N \;, 
\end{equation}
where we used \eqref{cond} and where ${\cal A}^N =v^M \, \nabla_M \, v^N$.
Since $v^N \, v_N = -1$ we have $v_N \, {\cal A}^N =0$, and hence
${\cal A}^N$ is proportional to the normal $n^N$, i.e.
${\cal A}^N = {\hat {\cal A}} \, n^N$.  To compute ${\hat {\cal A}}$
we use the fact that the black hole metric \eqref{bhline}
has a timelike Killing vector $l = l^M \, \partial_M = \partial_t$.
We compute $\partial_{\tau} \left( l_M \, v^M \right) = v^N \, \nabla_N 
\left( l_M \, v^M \right) = l_N \, {\cal A}^N = l_N \, n^N \, {\hat {\cal A}}
$, where we used the Killing
equation $\nabla_M \, l_N = - \nabla_N \, l_M$ once.  Hence
\cite{Visser:1989kg,Chamblin:1999ya}
\begin{equation}
K_{\tau \tau} = - \frac{\partial_{\tau} \left(l^M \, v_M \right)}{l^N \, n_N}
= - \frac{\partial_{\tau} v_t}{n_t} \;.
\end{equation}
Thus, the $\tau \tau$-component of \eqref{kmntracek} yields
\begin{equation}
\partial_{\tau} v_t = \frac{ \mathfrak{g}}{3} \, W \, n_t \;.
\label{eqhdot}
\end{equation}
Using \eqref{relationAB}, \eqref{dtdta} and \eqref{hubble}, we find that
\eqref{eqhdot} results in 
\begin{equation}
{\dot H} + H^2 =  \left( \frac{\mathfrak{g}}{3} \, W \right)^2
\left( 1 + \frac{a}{W} \, \frac{\partial W}{d a} \right) + \frac{1}{2 B^2 \, a}
\frac{\partial B}{\partial a} \;,
\end{equation}
where ${\dot H} = d H / d \tau$.  This equation is precisely the 
$\tau$-derivative of \eqref{hubble}.

Finally, we compute $K_{\tau i} = K_{M i} \, v^M$ using \eqref{defKMN}
and find $K_{\tau i} =0$, which is consistent with \eqref{kmntracek}.
Thus, we conclude that the equations \eqref{kmntracek} consistently
reduce to the Friedmann equation \eqref{hubble} and its $\tau$-derivative.

The Friedmann equation \eqref{hubble}
can be rewritten in terms of the black hole data $M$ and $Q_A \, \phi^A$
as follows.  Using \eqref{rarel} as well as
\eqref{relqUW} we obtain
\begin{equation}
{\rm e}^{2 U} = \frac{W}{3 \, a^2 - q_A \, X^A} \; a^2 \;.
\label{eua}
\end{equation}
Using \eqref{eua}, $Q_A \, \phi^A$ given in \eqref{qphi} becomes
\begin{eqnarray}\label{Qphi2}
 Q_A \, \phi^A =  \frac{2}{3 w_5} 
\left(-3k \,\frac{q_AX^A}{W}+k \, \frac{ \left(q_AX^A\right) \,
\left(q_B X^B\right)
}{Wa^2}
+\mu \, \frac{q_A X^A}{a^2}+k \, q_A h^A \right) \;.
\end{eqnarray}
Inserting \eqref{fgh}, \eqref{floweqw} and \eqref{eua} 
into \eqref{hubble} and using \eqref{Qphi2} yields 
\begin{equation}
 H^2=-k \, 
\frac{W \,\tilde W}{3 \,a^2}+\frac{w_5W}{3 \, a^4}M
-\frac{w_5 \,W}{6 \,a^4}\, Q_A \, \phi^A \;.
\label{friedblack}
\end{equation}

It is instructive to check whether $H^2 \geq 0$ along the motion of the
brane in the AdS black hole background.  Let us therefor consider
the black hole background \eqref{solfreedp23} for concreteness.  We set
$\mu=q=1$ (in appropriate units). 
The existence of a horizon then requires taking
$\mathfrak{g} < 1$, as shown in figure \ref{fig:r2g}.
\begin{figure}[t]
\centering
\includegraphics[bb=91 3 322 146]{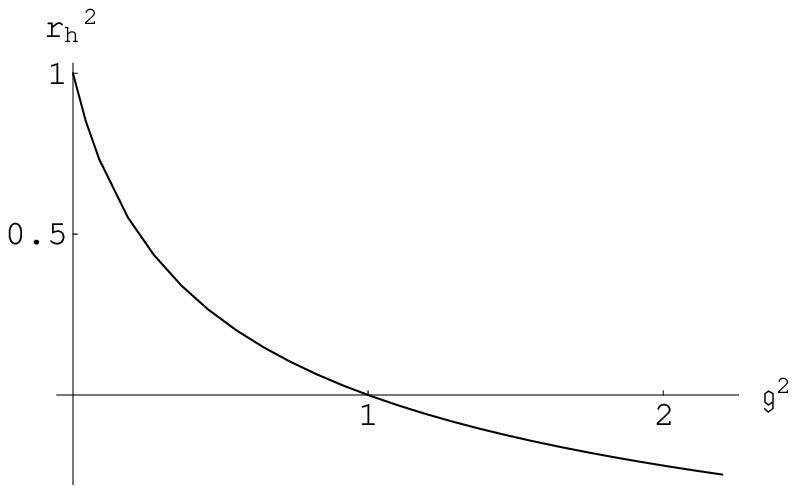}
\caption{$r^2_h$ over $\mathfrak{g}^2$. The physical range is 
$\mathfrak{g} < 1$. }
\label{fig:r2g}
\end{figure}
Picking the value $\mathfrak{g}^2 = 0.30$
we find that the horizon is at $r_h=0.64$
and that $H^2$ vanishes
at $r=1$. In 
figure \ref{fig:h2r} $H^2$ is plotted over $r$
using \eqref{friedblack}.
We find that  $H^2 \geq 0$ in the outside region between the horizon and
the turning point at $r=1$. 
The brane thus expands until it reaches its maximal
radius at $r=1$ after which it recontracts and falls through the horizon.
\begin{figure}[t]
\centering
\includegraphics[bb=91 3 322 146]{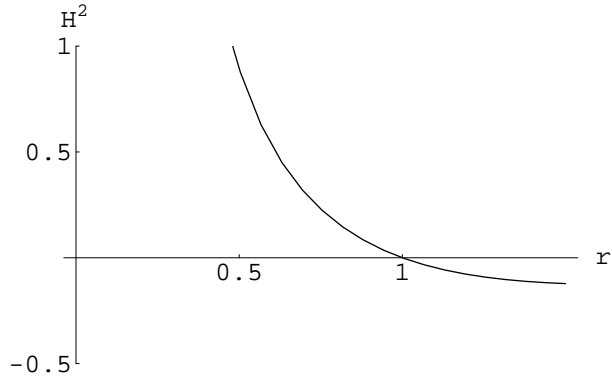}
\caption{$H^2$ over $r$.  The physical range is $r \leq 1$. }
\label{fig:h2r}
\end{figure}

\subsection{Dual description on the brane}

The AdS black hole provides a dual description of the radiation CFT
on the brane at finite temperature \cite{Witten:1998zw}.
Here the mass $M$, the Hawking temperature
$T_H$ and the entropy ${\cal S}$ of the black hole
are related
to the energy $E$ of the radiation on the brane, to its temperature $T$ and
to its entropy ${\cal S}$.  
The relation makes use of a 
conversion factor which is determined by the
asymptotic behaviour of $dt/d \tau = A^{-1/2}$
\cite{Savonije:2001nd}.  
For the line element
\eqref{bhline} it is given by $\left( {\mathfrak{g} \,a } \right)^{-1} $.
In the charged case, 
the black hole and field theory data are thus related as in table \ref{dict}
\cite{Savonije:2001nd,Biswas:2001sh,Youm:2001qr},
where $Q_A \, \phi^A_h$ denotes the horizon value of \eqref{qphi}
and where we recall that $L =  \mathfrak{g}^{-1}$.

\begin{table}[t]
\begin{center}
 \begin{tabular}{|c|c|}
 \hline
   \;\;\;\;\;\;\;\;AdS\;\;\;\;\;\;\;\; & \;\;\;\;\;\;\;\;CFT\;\;\;\;\;\;\;\;\
\\
 \hline
 \hline
   $M$ & $E=M \, L/a$\\
   $T_H$ & $T=T_H \, L/a$\\
   $Q_A \, \phi^A_h$ & $Q_A \, {\hat \Phi}^A= Q_A \, \phi^A_h \,L/a$\\
   $Q_A \, \phi^A$ & $Q_A \, \Phi^A= Q_A \, \phi^A \,L/a$\\
   $\cal{S}$ & $\cal{S}$ \\
 \hline
 \end{tabular} 
\caption{Relation between black hole and field theory data.
 \label{dict}}
\end{center}
\end{table}
\noindent

The spatial volume of the brane is given by
\begin{equation}
V = a^3 \, {vol} \;,
\label{vol}
\end{equation}
with ${vol}$ described below \eqref{defw5}.
The energy density $\rho$ of the radiation is
\begin{equation}
\rho = \frac{E}{V}
\label{rhobrane}
\end{equation}
and its pressure satisfies $p = \frac13 \rho$.
The energy $E$ is not a purely extensive quantity.  It contains a
sub-extensive part called the Casimir energy defined  
by \cite{Verlinde:2000wg}
\begin{equation}
E_c = 3 \left( E + p \, V - T \, {\cal S} - Q_A \, {\hat \Phi}^A \right) \;.
\label{casimir}
\end{equation}
Here we have defined $E_c$ in terms of ${\hat \Phi}^A$ rather than  
$\Phi^A$.
The Casimir energy \eqref{casimir} denotes the violation of the 
thermodynamic Euler relation.  The Euler relation for a 
system based on the first law of thermodynamics 
$dE = T \,d {\cal S} - p \,dV + 
{\hat \Phi}^A \,dQ_A$ states that if  
the energy $E({\cal S}, V, Q)$ is extensive, i.e. if it satisfies
$E(\lambda \,S, \lambda \,V, \lambda \,Q) = \lambda \, E( {\cal S}, V, Q)$,
then the energy takes the form 
$E = T \,{\cal S} - p \,V + Q_A \, {\hat \Phi}^A $. This relation is derived
by differentiating once with respect to $\lambda$, then setting
$\lambda =1$ and subsequently using the first law of thermodynamics.

The gravitational counterpart of the quantity $E_e$ 
and of the Casimir energy $E_c$ is given in \eqref{2mEcw} and \eqref{tildeEcw},
respectively. 
Using table \ref{dict},  we obtain
\begin{eqnarray}\label{EeEcW}
E_e &=& \frac{1}{\mathfrak{g} \, a} {\tilde E}_e = 
\frac{1}{4 \pi} \left(\frac{2 \,G_4}{vol}\right)^{1/3} \, 
\mathfrak{g}^{2/3} \, W_h \, \frac{{\cal S}^{4/3}}{a}  
\;, \nonumber\\
E_c &=& \frac{1}{\mathfrak{g} \, a} {\tilde E}_c =
\frac{3 k}{2 \pi} \left(\frac{vol}{2 \,G_4}\right)^{1/3} \, 
\frac{{\tilde W}_h}{\mathfrak{g}^{2/3}} 
\, \frac{{\cal S}^{2/3}}{a}  
 \;,
\end{eqnarray}
where we employed 
\eqref{g5g4} to rewrite \eqref{defw5} in terms of $G_4$. Observe that
$W_h$ and ${\tilde W}_h$
do not have a simple scaling behaviour under 
$a \rightarrow \lambda^{1/3} \, a \,,\,
{\cal S} \rightarrow \lambda \, {\cal S}, Q \rightarrow \lambda \, Q$.

Then, it follows that the entropy relation
\eqref{cvbh} can be written as 
\begin{equation}
k \, W_h \, {\tilde W}_h \, {\cal S}^2 = 
\frac{8 \pi^2}{3} \, a^2 \, E_c \, E_e =
\frac{4 \pi^2}{3} \, a^2 \, E_c \left(
2 E - Q_A \, {\hat \Phi}^A - E_c \right)  \
\label{cvcft}
\end{equation}
in terms of field theory data on the brane.
This is of the type of a Cardy-Verlinde formula 
(the higher-dimensional 
analogue of Cardy's formula ${\cal S} = 2 \pi \sqrt{
\frac{c}{6} \left(L_0 - \frac{c}{24} \right) }$ 
for the entropy of a two-dimensional CFT \cite{Cardy:1986ie}).

In \cite{Verlinde:2000wg} it was proposed that the entropy ${\cal S}$
of a CFT with an AdS-dual description
satisfies the bound ${\cal S} \leq {\cal S}_B$, where 
${\cal S}_B =\frac23 \pi \, a \, E $ denotes the Bekenstein entropy
\cite{Bekenstein:1980jp}.
To analyze whether this also holds 
for the expression
\eqref{cvcft} is not straightforward 
due to the presence of the factor $ W_h \, {\tilde W}_h$, which
satisfies the relation \eqref{wtilw3}.
For the
specific STU models \eqref{solfreedp23} and \eqref{free2}, however,
it is possible to check that the bound holds. For these models we have
$ 3 \, W \, {\tilde W} = 5 + 2 H^{-1}_1 + 2 H_1
= 9 + 2 q^2/(r^4 \, H_1)> 9$, since $H_1 >0$.  
It follows that 
$ W_h \, {\tilde W}_h > 3$.  {From} \eqref{solfreedp23} and \eqref{free2}
we see that 
the quantity $Q_A \, {\hat \Phi}^A $ is positive.  Therefore it follows that 
${\cal S} < \frac23 \pi \, a \sqrt{E_c \left(2E - E_c \right) }$,
which has a maximum at $E=E_c$ for a given energy $E$ \cite{Verlinde:2000wg},
and hence the bound ${\cal S} \leq {\cal S}_B$ is satisfied for these
two models.

Now we write the 
Friedmann equation \eqref{friedblack} in terms of 
field theory data (cf. table \ref{dict}).  Introducing the charge density
\begin{equation}
\rho_A = \frac{Q_A}{V} 
\end{equation}
we obtain
\begin{eqnarray}
 H^2=-k \,\frac{W \, \tilde W}{3 \,a^2}+\frac{8\pi \, G_4}{9} \, W \,
\left(\rho-\frac{1}{2}\rho_A \,\Phi^A\right) \;,
\label{friedcft}
\end{eqnarray}
where we used the relation \eqref{g5g4} to express the Friedmann equation
in terms of four-dimensional quantities.  
Observe that for any model \eqref{vsgconstraint}, 
$W$, ${\tilde W}$ and $\Phi^A$ are generically complicated functions of $a$.
Defining
\begin{eqnarray}\label{rhotot}
 \rho_{\rm eff}=\frac{W}{3}\rho-\frac{W}{6}
\, \rho_A \, \Phi^A+ k \, \frac{(3-W \, \tilde W)}{8\pi \,G_4 \, a^2} \;,
\end{eqnarray}
yields the Friedmann equation in the usual form, 
\begin{eqnarray}
 H^2=- \frac{k}{a^2}+\frac{8\pi \, G_4}{3}\rho_{\rm eff} \;.
\end{eqnarray}
This is the form following from \eqref{inteq} with 
a suitably chosen $T_{\mu \nu}^{\rm matter}$.

Next, we differentiate the Friedmann equation \eqref{friedcft} with respect to 
proper time $\tau$ to obtain 
\begin{eqnarray}
{\dot H} = \frac{ d H}{d \tau} &=& k \, \frac{W \, \tilde W}{3 \,a^2}
-\frac{4\pi \, G_4 W}{3} \left(\rho+p \right)
-\frac{k}{6 \, a} \, \frac{d}{da} \left(W\tilde W \right) 
\nonumber\\
&& +\frac{4\pi \, G_4 \, a}{9}\left(\frac{dW}{d a} \, 
\rho
-\frac{1}{2} \, \frac{d}{d a} \left(W \,
\rho_A \, \Phi^A \right )\right) \;,
\label{hdot}
\end{eqnarray}
where we used that radiation satisfies the equation
\begin{equation}
\frac{d \rho}{d \tau} = - 3 H \, \left(\rho + p \right) \;.
\end{equation}
The equation for $\dot H$ can be written in the standard FRW form
following from \eqref{inteq}, 
\begin{eqnarray}
 \dot H =\frac{k}{a^2}-4\pi \, G_4 \, \left(\rho_{\rm eff}+p_{\rm eff} \right) 
\;,
\end{eqnarray}
with $\rho_{\rm eff}$ given by \eqref{rhotot} and with $p_{\rm eff}$
given by 
\begin{eqnarray}
 p_{\rm eff}&=&\frac{W}{3}p+\frac{W}{6}\rho_A \, \Phi^A+\frac{k}{24\pi \, G_4
\, a^2}\left(a \, \frac{d}{da}
\left(W\tilde W \right) + W\tilde W-3\right) \nonumber\\
 & &-\frac{a}{9}
\left(\frac{dW}{d a} \, 
\rho
-\frac{1}{2} \,\frac{d}{d a} \left(W \,
\rho_A \, \Phi^A \right )\right) \;.
\end{eqnarray} 
Observe that $T_{\mu \nu}^{\rm matter}$ 
is, in general, not traceless,
\begin{eqnarray}
 T_{\mu}{}^{ \mu \; {\rm matter}}=-\rho_{\rm eff}+3p_{\rm eff}
&=&\frac{2}{3} \, W \, \rho_A \, \Phi^A+
k \, \frac{(W\tilde W-3 )}{4\pi \, G_4 \, a^2}
+\frac{k}{8\pi \, G_4 \,a}\frac{d}{da} \left(W\tilde W \right)  \nonumber\\
& &-\frac{a}{3}
\left(\frac{dW}{d a} \, 
\rho
-\frac{1}{2} \, \frac{d}{d a} \left(W \,
\rho_A \, \Phi^A \right )\right) \;.
\label{tracetmn}
\end{eqnarray}
In general, the energy density  $\rho_{\rm eff}$ is not of the
standard form.  Therefore 
we now check that $\rho_{\rm eff} > 0$
in the example considered before describing the motion of 
the brane in 
the black hole background \eqref{solfreedp23} with the parameters $\mu, q$
set to the values 
$\mu=q=1$. 
We find that $\rho_{\rm eff} > 0$
for $r >0$, as shown in figure \ref{fig:rtot}.
\begin{figure}[t]
\centering
\includegraphics[bb=91 3 322 146]{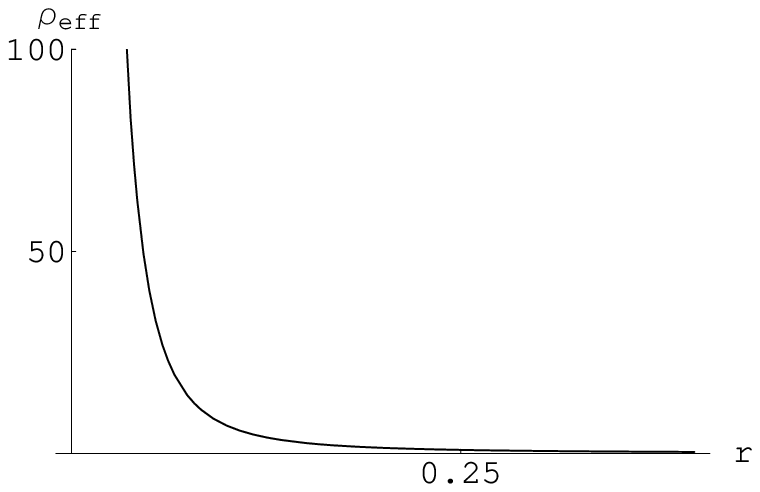}
\caption{$\rho_{\rm eff}$ over $r$.}
\label{fig:rtot}
\end{figure}
The behaviour of the pressure $p_{\rm eff}$ 
for the same values of the parameters
is displayed in figure 
\ref{fig:ptotout}.  It is positive throughout.
We also find that outside of the horizon, 
the trace of the energy-momentum tensor
$a^4 \, 
T_{\mu}{}^{ \mu \; {\rm matter}}$ only becomes vanishing asymptotically.
This is displayed in figure \ref{fig:a4tmm}.
\begin{figure}[t]
\centering
\includegraphics[bb=91 3 322 146]{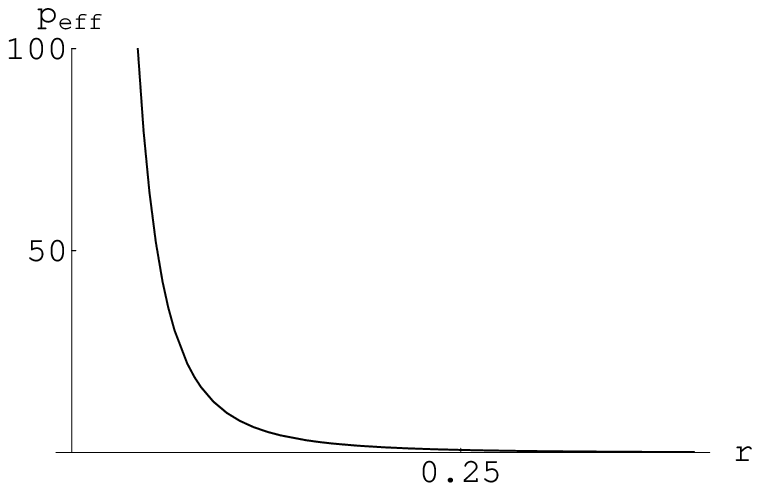}
\caption{$p_{\rm eff}$ over $r$.}
\label{fig:ptotout}
\end{figure}
\begin{figure}[t]
\centering
\includegraphics[bb=91 3 322 146]{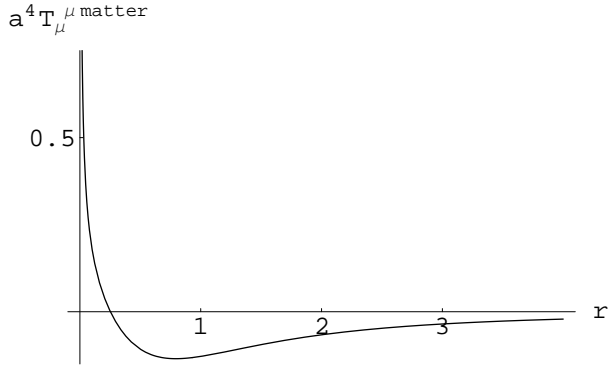}
\caption{$a^4 \, 
T_{\mu}{}^{ \mu \; {\rm matter}} $ over $r$.}
\label{fig:a4tmm}
\end{figure}

\subsection{Correspondence between the first FRW equation and the entropy}

The Friedmann equation \eqref{friedcft}, when written as
\begin{eqnarray}
k \, W \, {\tilde W} \, \left(\frac{9 \, V^2}{4 \, G_4^2 \, W^2} \,
 H^2 \right)= \frac{4 \pi^2}{3} \, a^2 \, \left(k \,
\frac{ 3  \, V \, {\tilde W} }{4 \pi \, G_4 \, a^2}
\right)
\, 
\left(2 E - Q_A \,\Phi^A
- k \frac{ 3 \, V \, {\tilde W} }{4 \pi \, G_4 \, a^2}
\right) \;,
\label{friedcftcv}
\end{eqnarray}
has a structure that is similar to that of the Cardy-Verlinde-type formula
\eqref{cvcft}.  When the brane crosses the event horizon, 
both equations coincide.
Namely, at the horizon where $f(r_h)=0$, \eqref{hubble} yields
\begin{equation} 
H^2_h = \left( \frac{\mathfrak{g}}{3} \, W_h \right)^2 \;,
\label{valuehubhor}
\end{equation}
which can be used to express $L =\mathfrak{g}^{-1} $ in terms of $H_h$
and $W_h$ \cite{Savonije:2001nd}.  Then, inserting this relation 
into the expression for the entropy \eqref{entro}
gives
\begin{equation}
{\cal S}^2  = \frac{9 \, V_h^2}{4 \, G_4^2 \, W^2_h} \, H^2_h \;,
\label{entrohorw}
\end{equation}
where $V_h$ denotes the volume \eqref{vol} evaluated at the horizon,
and where we used the relation \eqref{g5g4}.  On the other hand,
at the horizon
the Casimir energy $E_c$ can be written as
\begin{equation}
E_{c,h} = k \, \frac{3 \,V_h}{4 \pi \, G_4} \, \frac{{ \tilde W}_h}{a^2_h}
\label{casihor}
\end{equation}
by virtue of \eqref{entrohorw}. 
Inserting the expressions \eqref{entrohorw}
and \eqref{casihor} into \eqref{friedcftcv}
then yields the Cardy-Verlinde-type formula \eqref{cvcft}.

\subsection{Correspondence between the second FRW equation and the Casimir energy}

Now we rewrite the second FRW equation \eqref{hdot} in order
to exhibit its similarity with equation \eqref{casimir}
for the Casimir energy $E_c$. 
Using the second equation of \eqref{solflowZ} and \eqref{eua}
we obtain
\begin{eqnarray}
{\tilde W} &=& {\rm e}^{-2U} + \frac{q_A \, h^A}{3 \, a^2} \;, \nonumber\\
W \, {\tilde W} &=& 3 - \frac{q_A \, X^A}{a^2} + \frac{W}{3} \, \frac{q_A \, 
h^A}{a^2} \;,
\label{relww}
\end{eqnarray}
while
using \eqref{qphi} and \eqref{eua} we get
\begin{equation}
\frac{d \left( \rho_A \, \Phi^A \right)}{ d a} = - \frac{6}{a} \, \rho_A
\, \Phi^A + \frac{1}{4 \pi \, G_4 \, a^5} \left( 2 k \, q_A \, h^A - 6 k \, 
\frac{q_A \, X^A}{W} -
\left( k \, - \frac{\mu}{a^2}
\, {\rm e}^{2U} \right)
\, {\rm e}^{-2U} \, a \, q_A \frac{d X^A}{d a} 
\right) \;.
\label{derrhophi}
\end{equation}
Inserting \eqref{relww} and \eqref{derrhophi} 
into \eqref{hdot} and rearranging the terms we find that the second FRW equation \eqref{hdot}
can be written as 
\begin{eqnarray}
 k \, \frac{3 V \, {\tilde W}}{4\pi \, G_4 \,a^2}=
3\left( E + p \, V - Q_A \, \Phi^A
+\frac{3V}{4\pi \, G_4 W}\left( \frac{1}{9 \, a} 
\left( k \, - \frac{\mu}{a^2}
\, {\rm e}^{2U} \right) {\rm e}^{-4U} \, W \frac{d W}{d a}+{\dot H} \right)\right)
\label{FRW2casi}
\end{eqnarray}
by virtue of the relation $H_A \, d X^A/d a =0$, which holds due
to the second equation of 
\eqref{solflowZ} and the very special geometry relation \eqref{vsgxdx}.

Now let us consider the Casimir energy $E_c$.  Following
\cite{Savonije:2001nd}, we first relate the brane
temperature $T$ to $\dot H$.  Using the first equation in
\eqref{floweqw}, the Hawking
temperature \eqref{thawk} can be written as
\begin{equation}
T_H = 
\frac{1}{12 \pi}  \left[\frac{d f}{da} \, {\rm e}^{-4U} \, W\right]_h \;.
\end{equation}
Then, taking the $\tau$-derivative of \eqref{hubble} and using that
at the horizon $f(r_h)=0$ as well as  \eqref{valuehubhor},
we obtain for the temperature on the brane, 
\begin{eqnarray}\label{tempbrhor}
 T =T_H \, 
\frac{L}{a}= 
\frac{a_h}{2\pi \, a}  
\left[\frac{|H| }{W} \,a\, \frac{d W}{d a} -
\frac{\dot H}{|H|}\right]_h \;.
\end{eqnarray}
Inserting the expressions \eqref{entrohorw}, \eqref{casihor} and
\eqref{tempbrhor}
into the defining relation
\eqref{casimir} yields the
equation for the Casimir energy in the form
\begin{eqnarray}
\label{HubAb1}
k \, \frac{3 V_h \, {\tilde W_h}}{4\pi \, G_4 \,a_h \, a}=
3\left( E+p \, V-Q_A \,\hat \Phi^A
+\frac{3V_h \, a_h}{4\pi \, G_4 W_h \, a}\left[-\frac{H^2 }{W} \,a\, \frac{d W}{d a} +
\dot H\right]_h \right) \;.
\end{eqnarray}
Comparing \eqref{FRW2casi} with \eqref{HubAb1} shows that both equations
have a similar structure.  They coincide at the event
horizon of the black hole since 
\begin{equation}
\left[-\frac{H^2}{W} \, a \, \frac{d W}{d a}\right]_h = \left[
 \frac{1}{9 \, a} 
\left( k \, - \frac{\mu}{a^2}
\, {\rm e}^{2U} \right) {\rm e}^{-4U} \, W \frac{d W}{d a}\right]_h \;,
\end{equation}
where we used \eqref{valuehubhor} as well as  $f(r_h) =0$.

\section{Final comments}

It would be interesting to extend the analysis at the two-derivative
level given above and to include higher-curvature
terms such as the Gauss-Bonnet combination with a scalar-field
dependent coupling function.
The presence of these terms leads to further sub-extensive 
contributions that modify the form of the Cardy-Verlinde-type formula, the
Casimir energy and the FRW equations. 

A simpler example, on which we will now comment, is
provided by a
static AdS-Schwarz\-schild black hole 
modified by the presence of 
a  Gauss-Bonnet interaction term.
The black hole solution is known in closed form
\cite{Boulware:1985wk,Cai:2001dz,Torii:2005xu}, and the Friedmann equation
for a spherical brane moving in this black hole background has been
derived in \cite{Davis:2002gn,Gravanis:2002wy,Brihaye:2008kh}.
The bulk action is given by
\begin{equation}
S = \frac{1}{16 \pi \, G_5} \int d^5 x \sqrt{-g} \left[ R - 2 \Lambda + 
\alpha \left( R^2 - 4 R^{MN} R_{MN} + R^{MNPQ} R_{MNPQ} \right) \right] \;,
\label{bulkacgb}
\end{equation}
where $\Lambda = - 6 \, \mathfrak{g}^2 $, 
which corresponds to setting $W=3$ in \eqref{potent}.
In the context of string theory, $\alpha$ is proportional
to $\alpha'$.  
The associated black hole solution with a spherical horizon 
has the line element \cite{Boulware:1985wk,Cai:2001dz,Torii:2005xu}
(we use the notation of \cite{Brihaye:2008kh})
\begin{equation}
\label{linebhgb}
ds^2_5 = - f dt^2 + f^{-1} dr^2 + r^2 \, d\Omega^2_3 \;\;\;,\;\;\;
f = 1 + \frac{r^2}{4 \alpha} \left( 1 - \sqrt{1 + 8 \alpha 
\left(\frac{\mu}{r^4} -  \mathfrak{g}^2 \right)} \right) \;,
\end{equation}
where $d \Omega^2_3$ denotes the line element of a unit three-sphere.
The horizon of the black hole is located at $f (r_h)=0$, which yields
\begin{equation}
\mu = \mathfrak{g}^2 \, r^4_h + r^2_h + 2 \alpha    \;.
\label{horgaussb}
\end{equation}
The mass of the black hole is $M = \mu/w_5$.
The Gauss-Bonnet
corrected entropy reads \cite{Cai:2001dz,Dutta:2006vs}
\begin{equation}
{\cal S} = \frac{4 \pi}{3 w_5} 
\left( r^3_h + 12 \alpha \, r_h \right) \;,
\label{entrogb}
\end{equation}
and the Hawking temperature $T_H$ is given by
\begin{equation}
T_H = \frac{1}{4 \pi} \, f'(r_h) = 
\frac{1}{2 \pi \, r_h} \left( 1 + 2 \, \frac{\mathfrak{g}^2 \, r^4_h - 2 \alpha}{r^2_h + 4 \alpha} \right) \;.
\label{relfpt}
\end{equation}
The relation \eqref{entrogb}
can be inverted to express the horizon radius $r_h$ in terms of
the entropy,
\begin{equation}
r_h (s) = \frac{-8 \,\alpha+2^{1/3}\left(s+
\sqrt{s^2+256 \,\alpha^3}\right)^{2/3}}{2^{2/3}
\left(s+\sqrt{s^2+256 \,\alpha^3}\right)^{1/3}} \;,
\label{rhsmod}
\end{equation}
where 
\begin{equation}
s = \frac{3 w_5}{4 \pi} \, {\cal S} \;.
\end{equation}
Using \eqref{horgaussb} and \eqref{rhsmod} the mass $M$ can be expressed as
a power series in $s$.  At quadratic order in $\alpha$, this gives 
\begin{equation}
M= \frac{1}{w_5} \left(  \mathfrak{g}^2 \, 
s^{4/3} + \left(1 - 16 \alpha \, \mathfrak{g}^2\right)  
s^{2/3} -6 \alpha \left(1 - 16 \alpha \, \mathfrak{g}^2\right)   
+ 16 \alpha^2 \, s^{-2/3} \right) \;,
\end{equation}
which shows that 
the Gauss-Bonnet term induces an infinite series of sub-extensive
corrections expressed in powers of $\alpha \, s^{-2/3}$.

As it 
was the case at the two-derivative level, the bulk action \eqref{bulkacgb}
needs to be supplemented by both boundary terms \cite{Myers:1987yn}
and counterterms. 
The latter are given by \cite{Brihaye:2008kh}
\begin{equation} 
- \frac{1 }{8 \pi \, G_5} \int_{\Sigma} d^4 x \sqrt{- \gamma}
\left( c_1 
+ \frac{c_2}{2} \, {\cal R}
 \right) \;,
\label{counterc1c2}
\end{equation}
where
\begin{equation}
c_1 = \frac{1 + 8 \alpha \, \mathfrak{g}^2 - \sqrt{1 - 8 \alpha \,
\mathfrak{g}^2}}{2 \sqrt{\alpha}\, \sqrt{1 - \sqrt{1 - 8 \alpha \,
\mathfrak{g}^2}}} \;\;\;,\;\;\;
c_2  = \frac{
\sqrt{\alpha} \left( 3 - 8 \alpha \, \mathfrak{g}^2 - 3 \sqrt{1 - 8 \alpha \,
\mathfrak{g}^2} \right)
}{\left (1 - \sqrt{1 - 8 \alpha \,
\mathfrak{g}^2} \right)^{3/2} } \;.
\label{c1c2}
\end{equation}
Inspection of \eqref{counterc1c2} shows that the 
five- and four-dimensional Newton's constants
are related by
\begin{equation}
G_5 = c_2 \, G_4 \;.
\label{G5G4gb}
\end{equation}
In the limit $\alpha \rightarrow 0$ one recovers both 
the counterterms in 
\eqref{combinedaction5} and \eqref{g5g4}.

The equation
of motion for the brane moving in the black hole background \eqref{linebhgb}
is expressed in terms of both the extrinsic
curvature tensor $K_{MN}$ and the Riemann tensor on the brane 
\cite{Brihaye:2008kh}.  As in the two-derivative case (see \eqref{Keq} and \eqref{inteq}), we separate
the term proportional to the Einstein tensor on the brane from 
the other terms.  
We take the induced metric on the brane to have the form of
a standard FRW metric \eqref{metricbrane} 
with scale factor $a(\tau) = r (\tau)$
and Hubble parameter $H = {\dot r}/r$, where ${\dot r} = d r / d \tau$.
The resulting Friedmann equation takes the form
\cite{Davis:2002gn,Gravanis:2002wy}
\begin{equation}
\left(H^2 + \frac{f}{r^2} \right) \left(3 + 8 \alpha \left(H^2 + 
\frac{1}{r^2} \right) + 4 \alpha \frac{(1-f)}{r^2} \right)^2 = c^2_1 \;.
\label{modhubblegb}
\end{equation}
It can be checked that there is no induced cosmological constant on the
brane.  Indeed, 
dropping all 
the terms that involve powers of 
$\mu/r^4$ and of $H^2 + 1/r^2$ in
\eqref{modhubblegb} yields a perfect cancellation of all the remaining
terms.

At quadratic order in $\alpha$, \eqref{modhubblegb} gives
\begin{equation}
\label{friedmanngbexp}
H^2 = - \frac{1}{r^2} +  \left( 1 - 4 \alpha \, 
\mathfrak{g}^2 + 8 \alpha^2 \, \mathfrak{g}^4 \right) \frac{\mu}{r^4}
-2 \alpha \left(1 - 20 \alpha \,\mathfrak{g}^2 \right) \,
\frac{\mu^2}{r^8} + 8 \alpha^2 \, \frac{\mu^3}{r^{12}} \;.
\end{equation}
The energy $E$ on the brane is related to the mass $M$ by a
conversion factor which is determined by the
asymptotic behaviour of $dt/d \tau= f^{-1/2}$
\cite{Savonije:2001nd}.  
For the line element \eqref{linebhgb} this yields
the conversion factor $(\mathfrak{g} \, r)^{-1} (1 - \alpha \, \mathfrak{g}^2
- \frac52 \alpha^2 \, \mathfrak{g}^4 )$
at quadratic order in $\alpha$.  Using this as well as \eqref{G5G4gb} the
Friedmann equation \eqref{friedmanngbexp} can be expressed in terms
of the energy density $\rho$ \eqref{rhobrane} on the brane,
\begin{equation}
\label{frgbrho}
H^2 = - \frac{1}{r^2} +  \frac{8 \pi \, G_4}{3} \, \rho 
-2 \alpha \left(1 - 16 \alpha \,\mathfrak{g}^2 \right) \,
\left(\frac{8 \pi \, G_4}{3} \, \rho \right)^2
+ 8 \alpha^2 \, \left(\frac{8 \pi \, G_4}{3} \, \rho \right)^3
 \;.
\end{equation}
Observe that the coefficient of the 
term linear in $\rho$ does not receive $\alpha$-corrections.
It would be interesting to show that the equations for the entropy on
the brane (the Cardy-Verlinde formula and 
the equation for the Casimir energy) have a structure
that is similar to 
\eqref{frgbrho} and its $\tau$-derivative.

\subsection*{Acknowledgements}
We would like to thank M. Haack, 
D. L\"ust, D. Tsimpis and A. Yarom
for valuable discussions. 
This work is supported by the European Union contract 
MRTN-CT-2004-005104
``Constituents, Fundamental Forces and Symmetries of the Universe''.

\phantomsection
\addcontentsline{toc}{section}{\numberline{}References}
\bibliographystyle{JHEP-3}
\bibliography{bibliography}

\providecommand{\href}[2]{#2}\begingroup\raggedright\begin{thebibliography}{10}

\bibitem{Verlinde:2000wg}
E.~P. Verlinde, {\it On the holographic principle in a radiation dominated
  universe},  \href{http://arXiv.org/abs/hep-th/0008140}{{\tt hep-th/0008140}}.

\bibitem{Witten:1998zw}
E.~Witten, {\it Anti-de {S}itter space, thermal phase transition, and
  confinement in gauge theories},  {\em Adv. Theor. Math. Phys.} {\bf 2} (1998)
  505--532 [\href{http://arXiv.org/abs/hep-th/9803131}{{\tt hep-th/9803131}}].

\bibitem{Savonije:2001nd}
I.~Savonije and E.~P. Verlinde, {\it {CFT} and entropy on the brane},  {\em
  Phys. Lett.} {\bf B507} (2001) 305--311
  [\href{http://arXiv.org/abs/hep-th/0102042}{{\tt hep-th/0102042}}].

\bibitem{Randall:1999ee}
L.~Randall and R.~Sundrum, {\it {A large mass hierarchy from a small extra
  dimension}},  {\em Phys. Rev. Lett.} {\bf 83} (1999) 3370--3373
  [\href{http://arXiv.org/abs/hep-ph/9905221}{{\tt hep-ph/9905221}}].

\bibitem{Randall:1999vf}
L.~Randall and R.~Sundrum, {\it {An alternative to compactification}},  {\em
  Phys. Rev. Lett.} {\bf 83} (1999) 4690--4693
  [\href{http://arXiv.org/abs/hep-th/9906064}{{\tt hep-th/9906064}}].

\bibitem{Gubser:1999vj}
S.~S. Gubser, {\it {AdS/CFT} and gravity},  {\em Phys. Rev.} {\bf D63} (2001)
  084017 [\href{http://arXiv.org/abs/hep-th/9912001}{{\tt hep-th/9912001}}].

\bibitem{Cardy:1986ie}
J.~L. Cardy, {\it Operator content of two-dimensional conformally invariant
  theories},  {\em Nucl. Phys.} {\bf B270} (1986) 186--204.

\bibitem{Gunaydin:1984ak}
M.~Gunaydin, G.~Sierra and P.~K. Townsend, {\it Gauging the {D = 5}
  {M}axwell-{E}instein supergravity theories: more on {J}ordan algebras},  {\em
  Nucl. Phys.} {\bf B253} (1985) 573.

\bibitem{Cai:2001jc}
R.-G. Cai, {\it The {C}ardy-{V}erlinde formula and {AdS} black holes},  {\em
  Phys. Rev.} {\bf D63} (2001) 124018
  [\href{http://arXiv.org/abs/hep-th/0102113}{{\tt hep-th/0102113}}].

\bibitem{Klemm:2001pn}
D.~Klemm, A.~C. Petkou, G.~Siopsis and D.~Zanon, {\it Universality and a
  generalized c-function in {CFT}s with {AdS} duals},  {\em Nucl. Phys.} {\bf
  B620} (2002) 519--536 [\href{http://arXiv.org/abs/hep-th/0104141}{{\tt
  hep-th/0104141}}].

\bibitem{Biswas:2001sh}
A.~K. Biswas and S.~Mukherji, {\it Holography and stiff-matter on the brane},
  {\em JHEP} {\bf 03} (2001) 046
  [\href{http://arXiv.org/abs/hep-th/0102138}{{\tt hep-th/0102138}}].

\bibitem{Cai:2001ur}
R.-G. Cai, Y.~S. Myung and N.~Ohta, {\it {Bekenstein bound, holography and
  brane cosmology in charged black hole background}},  {\em Class. Quant.
  Grav.} {\bf 18} (2001) 5429--5440
  [\href{http://arXiv.org/abs/hep-th/0105070}{{\tt hep-th/0105070}}].

\bibitem{Youm:2001qr}
D.~Youm, {\it The {C}ardy-{V}erlinde formula and charged topological {AdS}
  black holes},  {\em Mod. Phys. Lett.} {\bf A16} (2001) 1327--1334
  [\href{http://arXiv.org/abs/hep-th/0105249}{{\tt hep-th/0105249}}].

\bibitem{Gregory:2002am}
J.~P. Gregory and A.~Padilla, {\it {Exact braneworld cosmology induced from
  bulk black holes}},  {\em Class. Quant. Grav.} {\bf 19} (2002) 4071--4083
  [\href{http://arXiv.org/abs/hep-th/0204218}{{\tt hep-th/0204218}}].

\bibitem{Bekenstein:1980jp}
J.~D. Bekenstein, {\it {A universal upper bound on the entropy to energy ratio
  for bounded systems}},  {\em Phys. Rev.} {\bf D23} (1981) 287.

\bibitem{Behrndt:1998jd}
K.~Behrndt, M.~Cvetic and W.~A. Sabra, {\it Non-extreme black holes of five
  dimensional {N = 2 AdS} supergravity},  {\em Nucl. Phys.} {\bf B553} (1999)
  317--332 [\href{http://arXiv.org/abs/hep-th/9810227}{{\tt hep-th/9810227}}].

\bibitem{Chamblin:1999tk}
A.~Chamblin, R.~Emparan, C.~V. Johnson and R.~C. Myers, {\it Charged {AdS}
  black holes and catastrophic holography},  {\em Phys. Rev.} {\bf D60} (1999)
  064018 [\href{http://arXiv.org/abs/hep-th/9902170}{{\tt hep-th/9902170}}].

\bibitem{Cvetic:1999ne}
M.~Cvetic and S.~S. Gubser, {\it Phases of {R}-charged black holes, spinning
  branes and strongly coupled gauge theories},  {\em JHEP} {\bf 04} (1999) 024
  [\href{http://arXiv.org/abs/hep-th/9902195}{{\tt hep-th/9902195}}].

\bibitem{Chamblin:1999hg}
A.~Chamblin, R.~Emparan, C.~V. Johnson and R.~C. Myers, {\it Holography,
  thermodynamics and fluctuations of charged {AdS} black holes},  {\em Phys.
  Rev.} {\bf D60} (1999) 104026
  [\href{http://arXiv.org/abs/hep-th/9904197}{{\tt hep-th/9904197}}].

\bibitem{Hawking:1999dp}
S.~W. Hawking and H.~S. Reall, {\it Charged and rotating {AdS} black holes and
  their {CFT} duals},  {\em Phys. Rev.} {\bf D61} (2000) 024014
  [\href{http://arXiv.org/abs/hep-th/9908109}{{\tt hep-th/9908109}}].

\bibitem{Gubser:2004xx}
S.~S. Gubser and J.~J. Heckman, {\it {Thermodynamics of {R}-charged black holes
  in $AdS_5$ from effective strings}},  {\em JHEP} {\bf 11} (2004) 052
  [\href{http://arXiv.org/abs/hep-th/0411001}{{\tt hep-th/0411001}}].

\bibitem{Gao:2004tv}
C.~J. Gao and S.~N. Zhang, {\it Higher dimensional dilaton black holes with
  cosmological constant},  {\em Phys. Lett.} {\bf B605} (2005) 185--189
  [\href{http://arXiv.org/abs/hep-th/0411105}{{\tt hep-th/0411105}}].

\bibitem{Elvang:2007ba}
H.~Elvang, D.~Z. Freedman and H.~Liu, {\it From fake supergravity to
  superstars},  \href{http://arXiv.org/abs/hep-th/0703201}{{\tt
  hep-th/0703201}}.

\bibitem{Astefanesei:2007vh}
D.~Astefanesei, H.~Nastase, H.~Yavartanoo and S.~Yun, {\it {Moduli flow and
  non-supersymmetric AdS attractors}},
  \href{http://arXiv.org/abs/arXiv:0711.0036 [hep-th]}{{\tt arXiv:0711.0036
  [hep-th]}}.

\bibitem{Lu:2003iv}
H.~Lu, C.~N. Pope and J.~F. Vazquez-Poritz, {\it From {AdS} black holes to
  supersymmetric flux-branes},  {\em Nucl. Phys.} {\bf B709} (2005) 47--68
  [\href{http://arXiv.org/abs/hep-th/0307001}{{\tt hep-th/0307001}}].

\bibitem{Miller:2006ay}
C.~M. Miller, K.~Schalm and E.~J. Weinberg, {\it Nonextremal black holes are
  {BPS}},  {\em Phys. Rev.} {\bf D76} (2007) 044001
  [\href{http://arXiv.org/abs/hep-th/0612308}{{\tt hep-th/0612308}}].

\bibitem{Janssen:2007rc}
B.~Janssen, P.~Smyth, T.~Van~Riet and B.~Vercnocke, {\it A first-order
  formalism for timelike and spacelike brane solutions},
  \href{http://arXiv.org/abs/arXiv:0712.2808 [hep-th]}{{\tt arXiv:0712.2808
  [hep-th]}}.

\bibitem{Gauntlett:1998fz}
J.~P. Gauntlett, R.~C. Myers and P.~K. Townsend, {\it Black holes of {D = 5}
  supergravity},  {\em Class. Quant. Grav.} {\bf 16} (1999) 1--21
  [\href{http://arXiv.org/abs/hep-th/9810204}{{\tt hep-th/9810204}}].

\bibitem{Balasubramanian:1999re}
V.~Balasubramanian and P.~Kraus, {\it A stress tensor for anti-de {S}itter
  gravity},  {\em Commun. Math. Phys.} {\bf 208} (1999) 413--428
  [\href{http://arXiv.org/abs/hep-th/9902121}{{\tt hep-th/9902121}}].

\bibitem{Emparan:1999pm}
R.~Emparan, C.~V. Johnson and R.~C. Myers, {\it Surface terms as counterterms
  in the {AdS/CFT} correspondence},  {\em Phys. Rev.} {\bf D60} (1999) 104001
  [\href{http://arXiv.org/abs/hep-th/9903238}{{\tt hep-th/9903238}}].

\bibitem{Kraus:1999it}
P.~Kraus, {\it Dynamics of anti-de {S}itter domain walls},  {\em JHEP} {\bf 12}
  (1999) 011 [\href{http://arXiv.org/abs/hep-th/9910149}{{\tt
  hep-th/9910149}}].

\bibitem{Kraus:1999di}
P.~Kraus, F.~Larsen and R.~Siebelink, {\it The gravitational action in
  asymptotically {AdS} and flat spacetimes},  {\em Nucl. Phys.} {\bf B563}
  (1999) 259--278 [\href{http://arXiv.org/abs/hep-th/9906127}{{\tt
  hep-th/9906127}}].

\bibitem{Batrachenko:2004fd}
A.~Batrachenko, J.~T. Liu, R.~McNees, W.~A. Sabra and W.~Y. Wen, {\it Black
  hole mass and {H}amilton-{J}acobi counterterms},  {\em JHEP} {\bf 05} (2005)
  034 [\href{http://arXiv.org/abs/hep-th/0408205}{{\tt hep-th/0408205}}].

\bibitem{Nojiri:2001ae}
S.~Nojiri, S.~D. Odintsov and S.~Ogushi, {\it Cosmological and black hole brane
  world universes in higher derivative gravity},  {\em Phys. Rev.} {\bf D65}
  (2002) 023521 [\href{http://arXiv.org/abs/hep-th/0108172}{{\tt
  hep-th/0108172}}].

\bibitem{Lidsey:2002zw}
J.~E. Lidsey, S.~Nojiri and S.~D. Odintsov, {\it {Braneworld cosmology in
  (anti)-de {S}itter {E}instein-{G}auss-{B}onnet-{M}axwell gravity}},  {\em
  JHEP} {\bf 06} (2002) 026 [\href{http://arXiv.org/abs/hep-th/0202198}{{\tt
  hep-th/0202198}}].

\bibitem{Nojiri:2002hz}
S.~Nojiri, S.~D. Odintsov and S.~Ogushi, {\it {Friedmann-Robertson-Walker brane
  cosmological equations from the five-dimensional bulk (A)dS black hole}},
  {\em Int. J. Mod. Phys.} {\bf A17} (2002) 4809--4870
  [\href{http://arXiv.org/abs/hep-th/0205187}{{\tt hep-th/0205187}}].

\bibitem{Cai:2002bn}
R.-G. Cai and Y.~S. Myung, {\it {Holography and entropy bounds in
  {G}auss-{B}onnet gravity}},  {\em Phys. Lett.} {\bf B559} (2003) 60--64
  [\href{http://arXiv.org/abs/hep-th/0210300}{{\tt hep-th/0210300}}].

\bibitem{Gregory:2003px}
J.~P. Gregory and A.~Padilla, {\it {Braneworld holography in Gauss-Bonnet
  gravity}},  {\em Class. Quant. Grav.} {\bf 20} (2003) 4221--4238
  [\href{http://arXiv.org/abs/hep-th/0304250}{{\tt hep-th/0304250}}].

\bibitem{Ferrara:1997tw}
S.~Ferrara, G.~W. Gibbons and R.~Kallosh, {\it {Black holes and critical points
  in moduli space}},  {\em Nucl. Phys.} {\bf B500} (1997) 75--93
  [\href{http://arXiv.org/abs/hep-th/9702103}{{\tt hep-th/9702103}}].

\bibitem{Sabra:1997yd}
W.~A. Sabra, {\it General {BPS} black holes in five dimensions},  {\em Mod.
  Phys. Lett.} {\bf A13} (1998) 239--251
  [\href{http://arXiv.org/abs/hep-th/9708103}{{\tt hep-th/9708103}}].

\bibitem{Larsen:2006xm}
F.~Larsen, {\it The attractor mechanism in five dimensions},
  \href{http://arXiv.org/abs/hep-th/0608191}{{\tt hep-th/0608191}}.

\bibitem{Girardello:1998pd}
L.~Girardello, M.~Petrini, M.~Porrati and A.~Zaffaroni, {\it Novel local {CFT}
  and exact results on perturbations of {N = 4} super {Y}ang-{M}ills from {AdS}
  dynamics},  {\em JHEP} {\bf 12} (1998) 022
  [\href{http://arXiv.org/abs/hep-th/9810126}{{\tt hep-th/9810126}}].

\bibitem{Freedman:1999gp}
D.~Z. Freedman, S.~S. Gubser, K.~Pilch and N.~P. Warner, {\it Renormalization
  group flows from holography supersymmetry and a c-theorem},  {\em Adv. Theor.
  Math. Phys.} {\bf 3} (1999) 363--417
  [\href{http://arXiv.org/abs/hep-th/9904017}{{\tt hep-th/9904017}}].

\bibitem{Behrndt:1999ay}
K.~Behrndt, {\it Domain walls of {D = 5} supergravity and fixpoints of {N = 1}
  super {Y}ang-{M}ills},  {\em Nucl. Phys.} {\bf B573} (2000) 127--148
  [\href{http://arXiv.org/abs/hep-th/9907070}{{\tt hep-th/9907070}}].

\bibitem{Behrndt:1999kz}
K.~Behrndt and M.~Cvetic, {\it Supersymmetric domain wall world from {D = 5}
  simple gauged supergravity},  {\em Phys. Lett.} {\bf B475} (2000) 253--260
  [\href{http://arXiv.org/abs/hep-th/9909058}{{\tt hep-th/9909058}}].

\bibitem{Skenderis:1999mm}
K.~Skenderis and P.~K. Townsend, {\it Gravitational stability and
  renormalization-group flow},  {\em Phys. Lett.} {\bf B468} (1999) 46--51
  [\href{http://arXiv.org/abs/hep-th/9909070}{{\tt hep-th/9909070}}].

\bibitem{Kallosh:2000tj}
R.~Kallosh and A.~D. Linde, {\it Supersymmetry and the brane world},  {\em
  JHEP} {\bf 02} (2000) 005 [\href{http://arXiv.org/abs/hep-th/0001071}{{\tt
  hep-th/0001071}}].

\bibitem{Behrndt:2000zh}
K.~Behrndt and S.~Gukov, {\it Domain walls and superpotentials from {M} theory
  on {C}alabi- {Y}au three-folds},  {\em Nucl. Phys.} {\bf B580} (2000)
  225--242 [\href{http://arXiv.org/abs/hep-th/0001082}{{\tt hep-th/0001082}}].

\bibitem{Behrndt:2000tr}
K.~Behrndt and M.~Cvetic, {\it Anti-de {S}itter vacua of gauged supergravities
  with 8 supercharges},  {\em Phys. Rev.} {\bf D61} (2000) 101901
  [\href{http://arXiv.org/abs/hep-th/0001159}{{\tt hep-th/0001159}}].

\bibitem{Gubser:2000nd}
S.~S. Gubser, {\it Curvature singularities: the good, the bad, and the naked},
  {\em Adv. Theor. Math. Phys.} {\bf 4} (2002) 679--745
  [\href{http://arXiv.org/abs/hep-th/0002160}{{\tt hep-th/0002160}}].

\bibitem{Ceresole:2001wi}
A.~Ceresole, G.~Dall'Agata, R.~Kallosh and A.~Van~Proeyen, {\it
  Hypermultiplets, domain walls and supersymmetric attractors},  {\em Phys.
  Rev.} {\bf D64} (2001) 104006
  [\href{http://arXiv.org/abs/hep-th/0104056}{{\tt hep-th/0104056}}].

\bibitem{Ceresole:2007wx}
A.~Ceresole and G.~Dall'Agata, {\it {Flow equations for non-BPS extremal black
  holes}},  {\em JHEP} {\bf 03} (2007) 110
  [\href{http://arXiv.org/abs/hep-th/0702088}{{\tt hep-th/0702088}}].

\bibitem{Maldacena:1997re}
J.~M. Maldacena, {\it The large {N} limit of superconformal field theories and
  supergravity},  {\em Adv. Theor. Math. Phys.} {\bf 2} (1998) 231--252
  [\href{http://arXiv.org/abs/hep-th/9711200}{{\tt hep-th/9711200}}].

\bibitem{Gubser:1998bc}
S.~S. Gubser, I.~R. Klebanov and A.~M. Polyakov, {\it Gauge theory correlators
  from non-critical string theory},  {\em Phys. Lett.} {\bf B428} (1998)
  105--114 [\href{http://arXiv.org/abs/hep-th/9802109}{{\tt hep-th/9802109}}].

\bibitem{Witten:1998qj}
E.~Witten, {\it Anti-de {S}itter space and holography},  {\em Adv. Theor. Math.
  Phys.} {\bf 2} (1998) 253--291
  [\href{http://arXiv.org/abs/hep-th/9802150}{{\tt hep-th/9802150}}].

\bibitem{Bianchi:2001kw}
M.~Bianchi, D.~Z. Freedman and K.~Skenderis, {\it Holographic renormalization},
   {\em Nucl. Phys.} {\bf B631} (2002) 159--194
  [\href{http://arXiv.org/abs/hep-th/0112119}{{\tt hep-th/0112119}}].

\bibitem{Smarr:1972kt}
L.~Smarr, {\it {Mass formula for Kerr black holes}},  {\em Phys. Rev. Lett.}
  {\bf 30} (1973) 71--73.

\bibitem{deHaro:2000wj}
S.~de~Haro, K.~Skenderis and S.~N. Solodukhin, {\it {Gravity in warped
  compactifications and the holographic stress tensor}},  {\em Class. Quant.
  Grav.} {\bf 18} (2001) 3171--3180
  [\href{http://arXiv.org/abs/hep-th/0011230}{{\tt hep-th/0011230}}].

\bibitem{Gibbons:1976ue}
G.~W. Gibbons and S.~W. Hawking, {\it Action integrals and partition functions
  in quantum gravity},  {\em Phys. Rev.} {\bf D15} (1977) 2752--2756.

\bibitem{Henningson:1998gx}
M.~Henningson and K.~Skenderis, {\it The holographic {W}eyl anomaly},  {\em
  JHEP} {\bf 07} (1998) 023 [\href{http://arXiv.org/abs/hep-th/9806087}{{\tt
  hep-th/9806087}}].

\bibitem{Burgess:1999vb}
C.~P. Burgess, N.~R. Constable and R.~C. Myers, {\it The free energy of {N = 4}
  super {Y}ang-{M}ills and the {AdS/CFT} correspondence},  {\em JHEP} {\bf 08}
  (1999) 017 [\href{http://arXiv.org/abs/hep-th/9907188}{{\tt
  hep-th/9907188}}].

\bibitem{Chamblin:1999ya}
H.~A. Chamblin and H.~S. Reall, {\it Dynamic dilatonic domain walls},  {\em
  Nucl. Phys.} {\bf B562} (1999) 133--157
  [\href{http://arXiv.org/abs/hep-th/9903225}{{\tt hep-th/9903225}}].

\bibitem{Brown:1992br}
J.~D. Brown and J.~York, James~W., {\it Quasilocal energy and conserved charges
  derived from the gravitational action},  {\em Phys. Rev.} {\bf D47} (1993)
  1407--1419.

\bibitem{Visser:1989kg}
M.~Visser, {\it Traversable wormholes from surgically modified {S}chwarzschild
  space-times},  {\em Nucl. Phys.} {\bf B328} (1989) 203.

\bibitem{Boulware:1985wk}
D.~G. Boulware and S.~Deser, {\it {String generated gravity models}},  {\em
  Phys. Rev. Lett.} {\bf 55} (1985) 2656.

\bibitem{Cai:2001dz}
R.-G. Cai, {\it Gauss-{B}onnet black holes in {AdS} spaces},  {\em Phys. Rev.}
  {\bf D65} (2002) 084014 [\href{http://arXiv.org/abs/hep-th/0109133}{{\tt
  hep-th/0109133}}].

\bibitem{Torii:2005xu}
T.~Torii and H.~Maeda, {\it {Spacetime structure of static solutions in
  {G}auss-{B}onnet gravity: neutral case}},  {\em Phys. Rev.} {\bf D71} (2005)
  124002 [\href{http://arXiv.org/abs/hep-th/0504127}{{\tt hep-th/0504127}}].

\bibitem{Davis:2002gn}
S.~C. Davis, {\it {Generalised {I}srael junction conditions for a
  {G}auss-{B}onnet brane world}},  {\em Phys. Rev.} {\bf D67} (2003) 024030
  [\href{http://arXiv.org/abs/hep-th/0208205}{{\tt hep-th/0208205}}].

\bibitem{Gravanis:2002wy}
E.~Gravanis and S.~Willison, {\it {Israel conditions for the {G}auss-{B}onnet
  theory and the {F}riedmann equation on the brane universe}},  {\em Phys.
  Lett.} {\bf B562} (2003) 118--126
  [\href{http://arXiv.org/abs/hep-th/0209076}{{\tt hep-th/0209076}}].

\bibitem{Brihaye:2008kh}
Y.~Brihaye and E.~Radu, {\it {Five-dimensional rotating black holes in
  {E}instein-{G}auss- {B}onnet theory}},
  \href{http://arXiv.org/abs/arXiv:0801.1021 [hep-th]}{{\tt arXiv:0801.1021
  [hep-th]}}.

\bibitem{Dutta:2006vs}
S.~Dutta and R.~Gopakumar, {\it {On {E}uclidean and {N}oetherian entropies in
  {AdS} space}},  {\em Phys. Rev.} {\bf D74} (2006) 044007
  [\href{http://arXiv.org/abs/hep-th/0604070}{{\tt hep-th/0604070}}].

\bibitem{Myers:1987yn}
R.~C. Myers, {\it {Higher derivative gravity, surface terms and string
  theory}},  {\em Phys. Rev.} {\bf D36} (1987) 392.

\end{thebibliography}\endgroup

\end{document}